\documentclass[aps,pre,amsmath,amssymb,twocolumn,showpacs,10pt]{revtex4-2}
\usepackage{graphicx}
\usepackage{dcolumn}
\usepackage{bm}
\usepackage{xcolor}
\usepackage{mathtools}
\usepackage{physics}
\usepackage{xcolor}
\usepackage[normalem]{ulem}
\usepackage{lineno}

\usepackage{url,hyperref}

\usepackage[english]{babel}
\usepackage{amsmath}
\usepackage{comment}
\usepackage{graphicx}
\usepackage{amssymb}

\newcommand \be  {\begin{equation}}
\newcommand \bea {\begin{eqnarray} \nonumber }
\newcommand \ee  {\end{equation}}
\newcommand \eea {\end{eqnarray}}

\newcommand{\corr}[1]{\langle#1\rangle}

\DeclareMathOperator{\diag}{diag}

\begin{document}

\title{ Intensity statistics inside an open wave-chaotic cavity with broken time-reversal invariance
}

\author{Yan V. Fyodorov}
\address{King's College London, Department of Mathematics, London  WC2R 2LS, United Kingdom}

\author{Elizaveta Safonova}
\address{Moscow Institute of Physics and Technology, Dolgoprudny, Russia}
\address{L. D. Landau Institute for Theoretical Physics, 142432 Chernogolovka, Russia}

\date{\today}

\begin{abstract}
Using the supersymmetric method of random matrix theory within the Heidelberg approach framework we provide statistical description of  stationary intensity sampled in locations inside an open wave-chaotic cavity, assuming that the time-reversal invariance inside the cavity is fully broken.
In particular, we show that when incoming waves are fed via a finite number $M$ of open channels the probability density ${\cal P}(I)$ for the single-point intensity $I$ decays as a power law for large intensities: ${\cal P}(I)\sim I^{-(M+2)}$, provided there is no internal losses. This behaviour is in marked difference with the Rayleigh law ${\cal P}(I)\sim \exp(-I/\overline{I})$ which turns out to be valid only in the limit $M\to \infty$. We also find the joint probability density of intensities $I_1, \ldots, I_L$  in $L>1$ observation points, and then extract the corresponding
 statistics for the maximal intensity in the observation pattern.  For $L\to \infty$ the resulting limiting  extreme value statistics (EVS)
 turns out to be different from the classical EVS distributions.
\end{abstract}

\maketitle

\section{Introduction}

This work aims to contribute towards understanding the statistics of intensity of a monochromatic wave field inside an
irregularly shaped enclosure (cavity) which could be fed with incoming waves through M open channels (antennae), see a sketch
in the figure:

\vspace{-2cm}

\begin{figure}[h!]
\includegraphics[scale=0.2,angle=00]{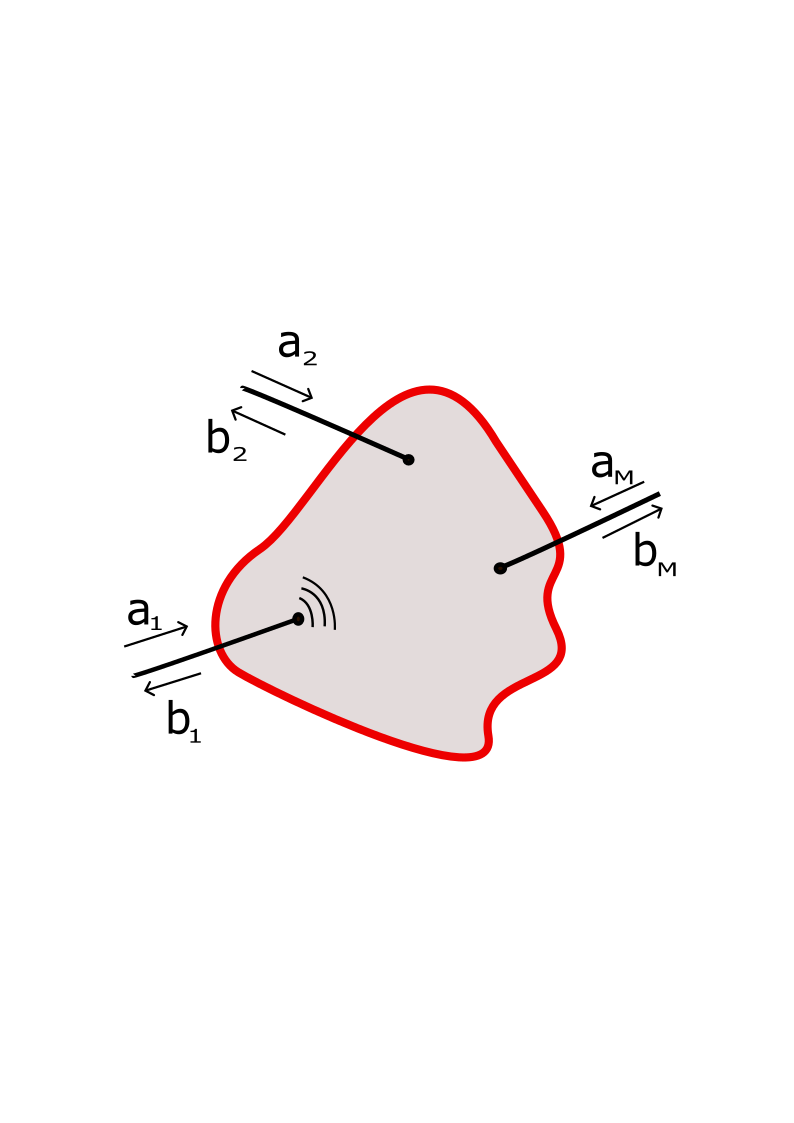}

\vspace{-2cm}

\hspace{7cm} $\left(\begin{array}{c}b_1\\ \ldots\ \\ \ldots \\ \ldots \\ b_M\end{array}\right)=\hat{S}\left(\begin{array}{c}a_1\\
\ldots\ \\ \ldots \\ \ldots \\ a_M\end{array}\right)$,  \hspace{1cm} $\psi_l=\frac{1}{\sqrt{k}}\left(a_l\,e^{-ikx_l}+b_l\,e^{ikx_l}\right)$.
\caption{\footnotesize A schematic sketch of a chaotic wave scattering in a cavity, with  $\psi_l, \, l=1,\ldots,M$ being the wave in $l_{th}$ channel/antenna,
with $x_l$ being the coordinate along the channel.  An operator describing wave dynamics in the system
decoupled from the channels/antennae is assumed to be effectively described by a large random matrix $\hat{H}$. The $M\times M$ unitary scattering matrix $\hat{S}$
can be related to $\hat{H}$ in the framework of the Heidelberg approach.}

\end{figure}

According to the standard paradigm of Quantum Chaos, we assume that the shape of the enclosure ensures chaotic ergodization
of a single classical particle motion in the same scattering domain.  At this ergodic situation universal
properties of closed wave-chaotic systems can be, following  the famous Bohigas-Giannoni-Schmidt (BGS) conjecture
\cite{BGS1984}, effectively modelled by replacing the microscopic system's Hamiltonian (or wave operator) by random matrices $\hat{H}$ of large dimension $N\gg 1$.
The standard choice is to use three ensembles with Gaussian-distributed entries, GOE, GUE and GSE, composed of real symmetric, complex Hermitian and real quanternionic matrices, respectively, and labelled by the Dyson parameter $\beta=1,2,4$. The choice $\beta=1$ is used to describe time-reversal invariant systems, $\beta=2$ coresponds to broken
time reversal symmetry, and $\beta=4$ to systems with Kramers degeneracy of energy levels.

The ensuing statistical characteristics of closed quantum-chaotic systems turn out to be universal, i.e. independent of
 microscopic details, when studied in the energy/frequency intervals of the length comparable to the typical distance $\Delta$ between neighbouring energy/frequency  levels.
It is expected that essentially the same statistics should be observed in regularly-shaped cavities
with a finite density of randomly placed scatterers inside, provided one neglects the effects of Anderson localization. The scale $\Delta$ is
assumed to be much smaller than the energy scale of the order of
inverse relaxation time $t_e$ ensuring full ergodization in the chaotic enclosure. In the context of systems with disorder
such time is controlled by classical diffusion and the corresponding energy scale is known as the Thouless time. Although proving BGS conjecture remains one of the great challenges in Mathematics, see e.g. \cite{RudnickGOE}, its validity at the level of Theoretical Physics is beyond any reasonable doubt, being supported by extensive numerics, as well as by elaborate field theory \cite{muzykantskii1995effective,AASA1996} and semiclassical
computations \cite{richter2002semiclassical,heusler2007periodic}.

Chaotic wave scattering in  enclosures is an  object of intensive research effort extending over several decades, with application to studies in
 compound nucleus scattering \cite{mitchell2010random},   transport properties in mesoscopic electronic systems \cite{Beenakker1997RevModPhys}
 and more recently in lasing  \cite{cao2015dielectric} as well as in manipulating light in complex media for energy deposition and imaging purposes \cite{rotter2017light}.
One of the central objects in both theory and experiments is the energy-dependent (or, rather, in the classical wave scattering context, frequency-dependent)
unitary scattering matrix (or simply $\hat{S}-$matrix)  $\hat{S}(E)$, the elements of which describe
the relationship between the vector ${\bf a}=(a_1,\ldots,a_M)$ of amplitudes of $M$ incoming waves in all open channels to the vector ${\bf b}=(b_1,\ldots,b_M)$ of the amplitudes of outgoing
waves. Since the scattering process is essentially random, the properties of $\hat{S}$-matrix must be
described using statistical language, i.e. probability distributions and correlation functions.  In developing this description the Random Matrix theory (RMT) plays
the central role.

The modern use of RMT for describing chaotic wave scattering statistically goes back to the seminal work of the Heidelberg group\cite{VWZ1985grassmann} who suggested
to model the scattering matrix elements in the form
\begin{equation}\label{VWZ}
S_{cc'}(E)=\delta_{cc'}-2i\sum_{x,y=1}^N W^*_{cx}\left[\frac{1}{E-\hat{{\cal H}}_{{\text{eff}}}}\right]_{xy}W_{yc'},
\end{equation}
where $\hat{{\cal{H}}}_{\text{eff}}= \hat{H}-i\sum_c \textbf{w}_c\otimes \textbf{w}_c^{\dagger}$, with $\hat{H}$ being $N\times N$ random matrix replacing true Hamiltonian of the closed cavity, and
the energy-independent vectors of coupling amplitudes  $\textbf{w}_c=(W_{c1},\ldots,W_{cN})$ relate $N$ inner states in the chosen basis to $M$ open
  channels. Without restricting generality, one can take vectors $\textbf{w}_c$ as fixed orthogonal satisfying
\begin{equation}\label{orthchan}
\textbf{w}_c^{\dagger}\textbf{w}_{c'}=\gamma_c\delta_{cc'}, \quad  \quad \gamma_c>0 \,\, \forall
c=1,\ldots, M.
\end{equation}

The orthogonality condition ensures that the ensemble-averaged scattering
matrix can be assumed to be diagonal
\be \label{diagS}\corr{\hat{S}(E)}=\diag
 (\corr{S_1(E)},\ldots,\corr{S_M(E)}),
 \ee
  where as $N\to \infty$ at fixed $M$ one finds that $\quad \corr{S_c(E)}=\frac{1-i\gamma_c\corr{G}}{1+i\gamma_c\corr{G}}$ and we introduced the mean value $\corr{G}:=\lim_{\eta\to 0} \corr {G({\bf r},{\bf r},E+i\eta)}=\corr{Re\, G}-i\pi\rho(E)$ for the  diagonal entry of the (retarded) Green’s function of the underlying closed cavity:
 $G({\bf r},{\bf r}',E+i\eta) := \left\langle{\bf r}\right|(E+i\eta-\hat{H})^{-1}\left|{\bf r}'\right\rangle$. This implies
 that $\rho(E)$ is the mean density of states in the cavity, which at the level of RMT  is given as $N\to \infty$ by the Wigner semicircle: $\rho(E)=\frac{1}{2\pi}\sqrt{4-E^2}, \, |E|<2$.  Writing $\corr{G}=|\corr{G}|e^{-i\alpha}$ and defining $\tilde{\gamma}_c:=\gamma_c |\corr{G}|$ one finds
 \be
  |\corr{S_c(E)}|^2=\frac{1-2\sin{\alpha}\,\tilde{\gamma}_c+\tilde{\gamma}_c^2}{1+2\sin{\alpha}\,\tilde{\gamma}_c+\tilde{\gamma}_c^2}
 \ee
 implying
 \be
  g_c=\frac{1+|\corr{S_c(E)}|^2}{1-|\corr{S_c(E)}|^2}=\frac{1}{2\sin{\alpha}}\left(\tilde{\gamma}_c+\frac{1}{\tilde{\gamma}_c}\right)\ge 1.
\ee
The set of parameters $g_c, \, c=1,\ldots,M$ provides the complete description of coupling of the medium to scattering channels in the universal regime, with
 the ``perfect coupling'' value $g_c=1$ (happening when $\sin{\alpha}=1$ and $\tilde{\gamma}_c=1$) corresponding to $|\corr{S_c}|=0$. The latter condition physically
 implies absence of short-time (also known as ``direct'') scattering processes at the channel $c$ entrance: all the incoming flux penetrates inside the medium and participates in formation of long-living resonant structures. This situation is thus most interesting from theoretical point of view, and is frequently described by most elegant formulas.
 In the RMT model the perfect coupling may occur only at the center of the spectrum $E=0$, where $\corr{G}=-i$, and we restrict our calculations henceforth to that point.
 Let us mention also the opposite limit $g_c\to \infty$ corresponding to the channel $c$ closed for incoming and outgoing waves.

The above-described choice for the model provides the most convenient framework for studying  statistics of the scattering matrix on small energy/frequency scales, comparable with separation $\Delta$ between neighbouring resonant frequencies/energy levels  of a closed system by
utilizing  the powerful supersymmetry approach developed earlier by Efetov \cite{EfetovBook} in the context of disordered electronic systems.
 Over the years it allowed to compute explicitly many statistical characteristics of the $\hat{S}-$matrix and other closely related objects, see e.g. \cite{Fyodorov1997JMP,dittes2000decay,fyodorov2005scattering,kumar2013distribution,nock2014distributions} and references therein.

Nowadays the model experimental setups to test the theoretical predictions based on Random Matrix Theory (RMT) are mainly
systems of classical waves (acoustic or electromagnetic) scattered from specially built resonators, shaped in the form of the so-called chaotic billiards or/and
with added scatterers inside, see e.g. \cite{Kuhl2005Rev,Kuhl2013microwave}.
Under appropriate conditions, the associated Helmholtz equation for the electric field strength is scalar and mathematically
identical to the two-dimensional Schr\"{o}dinger equation of
a particle elastically reflected by the contour of the microwave resonator, i.e., of a quantum billiard.
 Alternatively,  experiments on chaotic wave scattering are performed on systems built with
microwave graphs, see e.g. \cite{hul2004experimental}.

Whereas a lot of efforts was devoted to study of transmission and reflection of waves, which pertains to measuring the wave field {\it outside} of the
scattering medium (or at its boundary with external world), an interesting question is also to understand the statistics of wave patterns {\it inside} the chaotic enclosure.
This question is especially natural in view of growing interest in various aspects of coherent manipulations of wave propagation in complex media for imaging, light storage,
electromagnetic compatibility tests etc, see e.g. \cite{cheng2014focusing,rotter2017light,durand2019optimizing,horodynski2020optimal,gros2020tuning,lin2023predicting} and references therein. The study of statistics of radiation intensity in random medium has a long history. In particular, it has been suggested to model the wave pattern as a random superposition of running  plane waves with complex coefficients \cite{pnini1996intensity,brouwer2003wave}:
\begin{equation}\label{Gaumod}
u({\bf r})=\sum_{\bf k}a({\bf k})e^{i{\bf kr}}
\end{equation}
where all wavevectors ${\bf k}$ have the same length, while
the amplitudes $a(\bf k)$ are chosen as random gaussian complex numbers. While in closed systems with preserved time-reversal invariance one has to assume
$a^*({\bf k}) =a (-{\bf k})$, the correlations between $a(\bf k)$ and $a(-\bf k)$ gradually diminish with increased degree of openness of the scattering system.  The simplest prediction of such a model was
 a one-parameter family of possible distributions for the point intensity $I=|u({\bf r})|^2$, reducing to the simple exponential/Rayleigh distribution
  ${\cal P}(I)\propto e^{-I/\overline{I}}$ for completely uncorrelated  $a(\bf k)$ and $a(-\bf k)$.
Despite favourably agreeing with some experimental results \cite{kim2005measurement}, the use of simple Gaussian model Eq.(\ref{Gaumod}) looks largely phenomenological, and certainly calls for a proper microscopic justification.

Motivated by this,  the present paper aims to investigate statistics of the intensity of wave field at a given point ${\bf r}$ inside a chaotic cavity relying on
the same assumptions as RMT-based  model Eq.(\ref{VWZ}) for the $\hat{S}-$matrix.  We will demonstrate that the framework of the Heidelberg approach gives a possibility to derive ${\cal P}(I)$ for any fixed number $M$ of open channels without further assumptions, at least in the
 simplest case of chaotic systems with fully broken time-reversal invariance. The latter is described at the level of RMT by $\hat{H}$ taken from the Gaussian Unitary Ensemble. We note that for such systems the eigenfunctions of the closed cavity are already complex, with independent, identically distributed complex and imaginary parts. Hence when applying the Gaussian wave ansatz Eq.(\ref{Gaumod}) to the open system with broken time-reversal invariance one would naturally expect that $a(\bf k)$ and $a(-\bf k)$ are  uncorrelated, implying the Rayleigh law  as the reference for the intensity distribution. We will indeed see how such a law emerges in the limit of very open system, with the number of scattering channels tending to infinity. However for any finite number of channels the ensuing distribution  ${\cal P}(I)$ of local intensity  is found to be very different and shows a powerlaw rather than exponential decay. As  scattering systems with broken time-reversal invariance are now routinely
realized both in "billiard"-type scattering experiments \cite{stoffregen1995microwave,so1995wave,chung2000measurement,dietz2009induced,dietz2019partial} and in chaotic scattering in microwave graphs, see e.g \cite{lawniczak2010experimental,chen2021TDres}, one may expect that the predicted behaviour may be eventually tested experimentally.

 \section{Formulation of the problem and the main results}
We recall that the incoming waves are fed into the cavity via $M$ channels $c=1,\ldots,M$, with amplitudes given by the vector ${\bf a}=(a_1,\ldots,a_M)$. This creates a
field inside the cavity which we think of as a vector ${\bf u}$ in the $N-$dimensional inner Hilbert space. In particular, for our purpose it is convenient to think of the position basis $\left|{\bf r}\right\rangle$, associated with an appropriate coordinate system inside the cavity domain, so that the quantities $u({\bf r})\equiv \left\langle {\bf r}|{\bf u}  \right\rangle$ give precisely the amplitude of the wave in a point ${\bf r}$ inside the cavity. The corresponding intensity is given by $I_{\bf r}=|u({\bf r})|^2$.

In the framework of the Heidelberg model  one can relate the vector ${\bf u}$ at a given value of the energy/frequency to the scattering matrix  as
( see e.g. \cite{dittes2000decay} or  Eq.(27) in \cite{Fyodorov1997JMP})
\begin{equation}\label{innpartS}
\boldsymbol{u}=\frac{1}{2}\frac{1}{E-\hat{H}}\hat{W}\left(\hat{\bf 1}_M+\hat{S}\right)\boldsymbol{a},
\end{equation}
where $\hat{\bf 1}_M$ stands for the identity matrix and $\hat{H}$ is the random matrix representing the inner Hamiltonian, while $\hat{W}$ is the matrix whose $M$ columns are channel vectors $\textbf{w}_c$. Further using an equivalent form of the scattering matrix given by
\begin{equation}\label{eq:S-matrix form1}
\hat{S}=\left(\hat{\bf 1}_M-i\hat{K}\right)\times\left(\hat{\bf 1}_M+i\hat{K}\right)^{-1},\quad\hat{K}=\hat{W}^{\dagger}\frac{1}{E-\hat{H}}\hat{W}
\end{equation}
 one can bring Eq.(\ref{innpartS}) to another well-known form, cf. e.g. Eq.(38) in \cite{sokolov1997simple},
conveniently written in the bra-ket notations as
\begin{equation}\label{innpartmain}
\left|\boldsymbol{u}\right\rangle= \frac{1}{E-\hat{H}+i\hat{W}\hat{W}^{\dagger}}\left|\textbf{w}_{\bf a}\right\rangle, \quad \left|\textbf{w}_{\bf a}\right\rangle \equiv \sum_{c=1}^M\,a_c\left|\textbf{w}_{c}\right\rangle
 \end{equation}
 and implying for the intensity a representation
\begin{equation}\label{intenstart}
I_{\bf r}=\left\langle{\bf r}\right| \frac{1}{E-\hat{H}+i\hat{W}\hat{W}^{\dagger}}\left|\textbf{w}_{\bf a}\right\rangle\left\langle\textbf{w}_{\bf a}\right| \frac{1}{E-\hat{H}-i\hat{W}\hat{W}^{\dagger}}\left|{\bf r}\right\rangle.
\end{equation}
This formula is the starting point of our calculation of  the probability density  ${\cal P}(I)$ for the single-point intensity $I=I_{\bf r}$.

 Relegating the technical part of the calculation, largely inspired by the methods of the works \cite{kumar2013distribution,nock2014distributions}, to the body of the paper, we start
with presenting and discussing our  main results below.

\subsection{ \small Single-point intensity distribution}
 Given the set of  coupling parameters $g_c\ge 1, c=1,\ldots,M$, define for a given $I>0$ the parameter  $\lambda_1>1$ as the (unique) solution of the equation
\begin{equation}\label{Igeneq}
I=\frac{\lambda_1-1}{2}\sum_{c=1}^M|a_c|^2\left(1-\frac{g_c-1}{\lambda_1+g_c}\right).
\end{equation}
 The existence and uniqueness of the solution follows from the fact that the right-hand side of Eq.(\ref{Igeneq}) is positive, monotonically increasing to infinity function of $\lambda_1$ in the whole range $\lambda_1\in[1,\infty)$, and is equal to zero when $\lambda_1=1$.
 The intensity distribution in a single spatial point is then characterized by the probability density given explicitly in terms of $\lambda_1$  as
 \begin{equation}\label{mainresultA}
 {\cal P}_M(I)=\frac{d}{dI}I\frac{d}{dI} {\cal F}_M(I), \quad \mbox{with}\,\, {\cal F}_M(I)=\sum_{c=1}^M|a_c|^2{\cal F}_c(I),
 \end{equation}
 with ${\cal F}_c(I)$ given by
 \[
 {\cal F}_c(I)=\frac{\lambda_1-1}{\left(2I+(\lambda_1-1)^2\sum\limits_{i=1}^M|a_{i}|^2\frac{g_{i}-1}{(\lambda_1+g_{i})^2}\right)\prod\limits_{j = 1}^M(\lambda_1+g_{j})}
 \]
 \begin{equation}\label{mainresult_general}
 \times \int_{-1}^{1}\,d\lambda_2\,
\, \frac{\lambda_2+\tilde{g}_c}{\lambda_1-\lambda_2}\prod_{k\ne c }^M(\lambda_2+g_{k})
 \end{equation}
where we introduced the notation:
 \begin{equation}\label{g1tilde}
  \tilde{g}_c=\frac{1+g_c\lambda_1}{g_c+\lambda_1}.
 \end{equation}

There are two special cases when the solution to Eq.(\ref{Igeneq}) can be explicitly written. The first one pertains to the situation when the incoming wave
is incident only via a single channel, which we can choose to correspond to the channels index $c=1$, whereas all other channels with $2\le c\le M$ may only support outgoing waves. Indeed, setting $a_1=1$ for simplicity, and $a_c=0,\, \forall c=2,\ldots,M$ we see that Eq.(\ref{Igeneq}) becomes quadratic and one immediately finds that
\begin{equation}\label{singleincomingA}
\lambda_1=I+\sqrt{1+2g_1I+I^2},
\end{equation}
which after some manipulations allows to show that
\begin{equation}\label{relations1}
 \quad \tilde{g}_1=-I+\sqrt{1+2g_1I+I^2} \quad
\end{equation}
and
\[
\frac{\lambda_1 + g_1}{\lambda_1-1}=\frac{I+1+\sqrt{1+2g_1I+I^2}}{2I},
\]
which further implies
\begin{equation}\label{relations2}
\frac{\lambda_1 + g_1}{\lambda_1-1}\left(2I+(\lambda_1-1)^2\frac{g_1-1}{(\lambda_1+g_1)^2}\right)=2\sqrt{1+2g_1I+I^2}.
\end{equation}
Correspondingly, Eq.(\ref{mainresult_general}) takes very explicit and rather elegant form
 \begin{equation}\label{mainresult_singlech}
 {\cal F}_M(I)=\frac{1}{2\sqrt{1+2g_1I+I^2}}\frac{1}{\prod_{j=2}^M(I+\sqrt{1+2g_1I+I^2}+g_j)}
 \end{equation}
 \[
 \times \int_{-1}^{1}\,d\lambda_2\,
\, \frac{\lambda_2-I+\sqrt{1+2g_1I+I^2}}{I+\sqrt{1+2g_1I+I^2}-\lambda_2}\prod_{k=2}^M(\lambda_2+g_k).
 \]
The remaining integral, hence the probability density for the intensity $I$,  can be evaluated in a closed form for any coupling strengths but general results are quite cumbersome.
In the simplest case of a single open channel one gets
\begin{equation}\label{singlechanold}
 {\cal P}_{M=1}(I)=\frac{1}{(1+2g_1I+I^2)^{3/2}}
\end{equation}
\[
\times \left(2g_1-3(g_1^2-1)\frac{I}{1+2g_1I+I^2}\right),
\]
whereas for the two-channel case the cumulative distribution of intensities is given by
  \[
\int_I^{\infty}{\cal P}_{M=2}(\tilde{I})\,d\tilde{I}=-\frac{g_2I(\lambda_1+g_1)}{(\lambda_1+g_2)^2(1+2g_1I+I^2)}
\]
\begin{equation}\label{twochan}
+\frac{1}{(1+2g_1I+I^2)^{1/2}}\left(1-\frac{2I(\lambda_1+g_1)}{(\lambda_1+g_2)^2}\right)
\end{equation}
\[
-\frac{1}{(1+2g_1I+I^2)^{3/2}}\frac{g_2I(I+g_1)}{\lambda_1+g_2},
\]
with $\lambda_1$ defined in Eq.(\ref{singleincomingA}).\\[0.5ex]

 The second special case corresponds to all scattering channels being of equal strength:   $g_c=g\ge 1$ for all $c=1,\ldots, M$. Defining the total incoming flux in all channels as
 \begin{equation}\label{influx}
 {\cal I}=\sum_{c=1}^M|a_c|^2
 \end{equation}
 and further introducing the ratio $J=I/{\cal I}$ one finds that
  \begin{equation}\label{allincoming}
\lambda_1=J+\sqrt{1+2g\,J+J^2},
\end{equation}
implying that again $\tilde{g}=-J+\sqrt{1+2gJ+J^2}$, and further finding
 \[
 {\cal F}_M(I)=\frac{1}{2\sqrt{1+2gJ+J^2}}
 \]
 \begin{equation}\label{mainresultequivchan}
 \times
  \int_{-1}^{1}
\, \frac{\lambda_2-J+\sqrt{1+2gJ+J^2}}{J+\sqrt{1+2gJ+J^2}-\lambda_2}
 \end{equation}
 \[
 \times \left(\frac{\lambda_2+g}{J+\sqrt{1+2g_1J+J^2}+g}\right)^{M-1}\,d\lambda_2\,.
 \]
 Evaluating the integral in the closed form, we may assume $M\ge 2$ as we already considered $M=1$ case above.
 We then find
 \[
 {\cal F}_M(I)=-\ln{\frac{\lambda_1-1}{\lambda_1+1}}-\frac{1}{\sqrt{1+2gJ+J^2}}
 \]
 \begin{equation}\label{mainresultequivchan1}
 +\sum_{p=0}^{M-2}\left(\begin{array}{c}M-1\\ p+1\end{array}\right)\frac{(-1)^p}{(\lambda_1+g)^{p+1}}f_p(I)\,,
 \end{equation}
 where we defined
 \begin{equation}
 f_p(I)=\frac{1}{2(p+2)}\frac{(\lambda_1+1)^{p+2}-(\lambda_1-1)^{p+2}}{\sqrt{1+2gJ+J^2}}
 \end{equation}
 \[
 -\frac{1}{(p+1)}\left((\lambda_1+1)^{p+1}-(\lambda_1-1)^{p+1}\right),
 \]
 with $\lambda_1$ defined in Eq.(\ref{allincoming}).

 The probability density ${\cal P}_M(I)$ is then obtained by substituting Eq.(\ref{mainresultequivchan1})
 into Eq.(\ref{mainresultA}). The most elegant result emerges if
 all channels are perfectly coupled, with $g=1$ implying $\lambda_1=2J+1$. After some algebra we get in that case
 \begin{equation}\label{mainresultequivchan2}
 {\cal F}_M(I)=-\ln{\frac{J}{J+1}}-\sum_{p=1}^{M}\left(\begin{array}{c}M\\ p\end{array}\right)\frac{(-1)^p}{p}\left[\left(\frac{J}{J+1}\right)^p-1\right],
 \end{equation}
 and after substituting  into Eq.(\ref{mainresultA}) the probability density for the intensity $I$ takes an especially simple form:
 \begin{equation}\label{perfect1}
 {\cal P}_M(I)=(M+1)\frac{{\cal I}^{M+1}}{(I+{\cal I})^{M+2}}.
 \end{equation}
 In fact for any coupling the tail behaviour can be easily extracted from Eqs. (\ref{mainresultA}) and (\ref{mainresultequivchan}) and has the same powerlaw form:
  setting $I\to \infty$ at a fixed value of $g$ one finds the tail ${\cal P}(I)\sim I^{-(M+2)}$.
We conclude that for any finite number of channels the ensuing powerlaw-tailed distribution is quite different from the  Rayleigh law predictions of
the "Gaussian random wave" model.  Note however that setting in Eq.(\ref{perfect1}) the number of channels to infinity in such a way that the incoming flux per channel remains finite: $\lim_{M\to \infty}{\cal I}/M=\overline{I}<\infty$ restores the Rayleigh law:
  \begin{equation}\label{Rayleigh}
 \lim_{M\to \infty}{\cal P}_M(I)=\frac{1}{\overline{I}}e^{-I/\overline{I}}.
 \end{equation}
 This fact supports the view that the Gaussian wave model is asymptotically accurate if
 scattering system is  open in an essentially semiclassic way, with many incoming channels supporting finite flux per channel.\\
  In the figure 2 we show the results for a direct numerical simulations of the Heidelberg model against our theoretical predictions, with a very satisfactory agreement between the two.

 \begin{figure}[h!]
  \includegraphics[scale=0.5,angle=00]{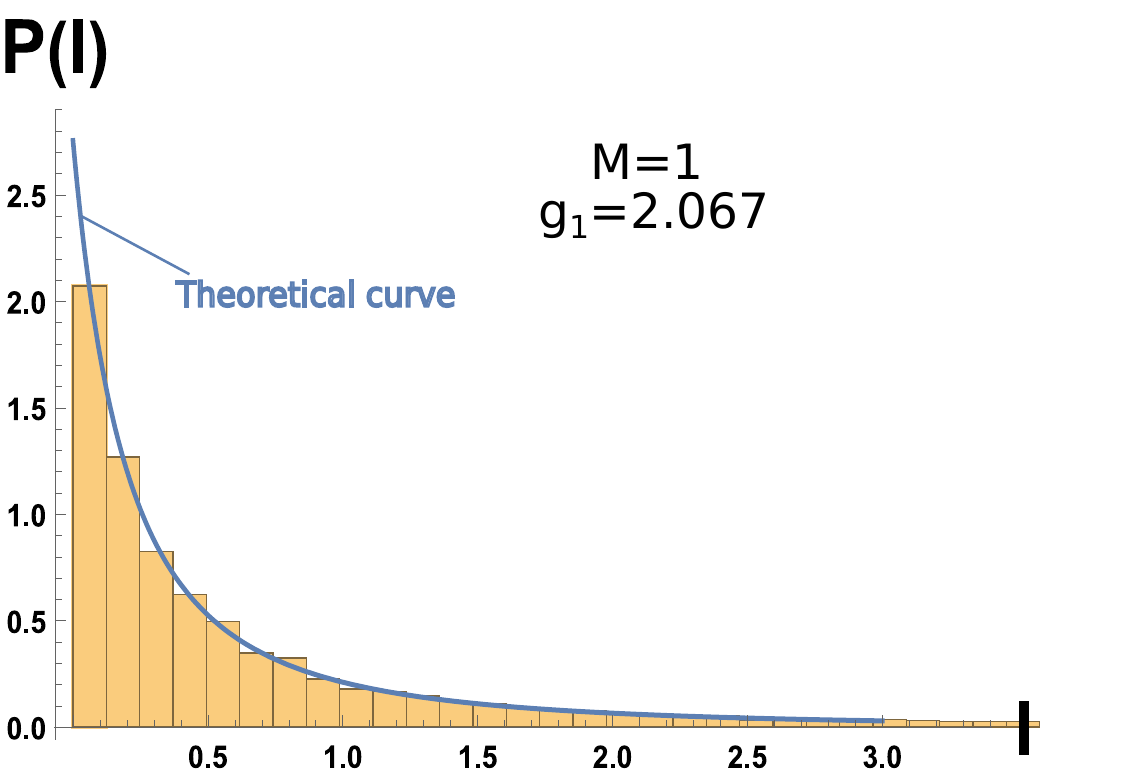}
\includegraphics[scale=0.5,angle=00]{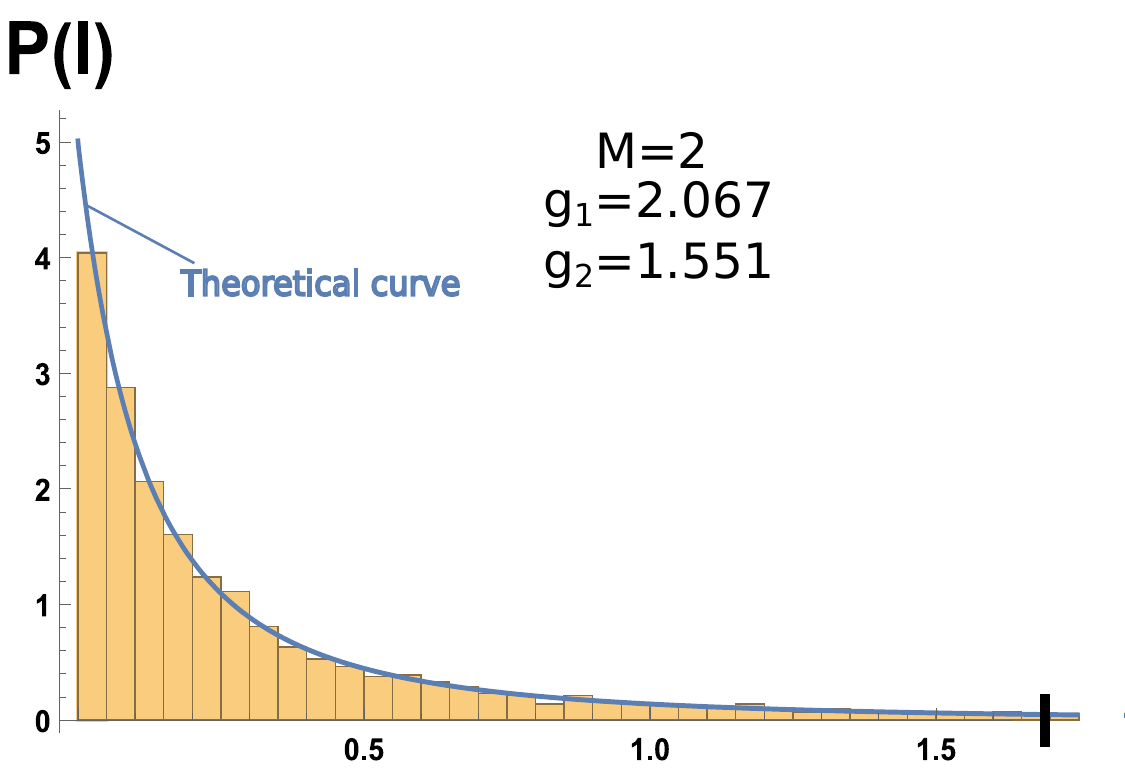}
\includegraphics[scale=0.5,angle=00]{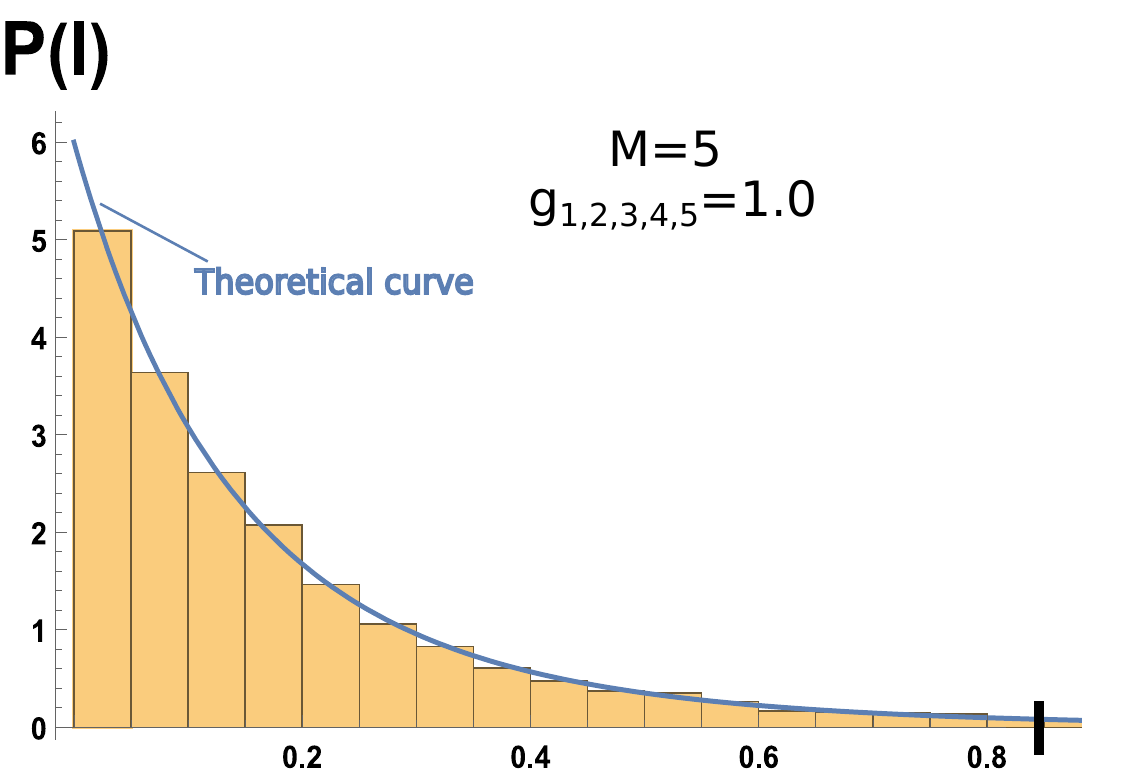}
\caption{\footnotesize The histogram shows numerically evaluated probability density $P(I)$ of the single-point intensity $I$
in the Heidelberg model for an M-channel system, with the incoming flux present only in the first channel. The intensity data  are generated according  to Eq.(\ref{intenstart})  in a given realization of GUE matrix of size $N=100$. Statistics is built using  10000 different GUE matrix realizations. The solid line is the theoretical prediction given by Eq.(\ref{singlechanold}) for $M=1$ with the coupling constant $g_1=2.067$, by  Eq.(\ref{twochan}) for $M=2$ and the coupling constants
 $g_1=2.067, \, g_2=1.551$ and Eq.(\ref{perfect1}) for $M=5$ perfectly coupled channels with $g_i=1$.}
\end{figure}

A few remarks are now in order.\\[1ex]

 {\bf Remark 1}. The  one-point intensity distribution presented in Eqs.(\ref{mainresultA})-(\ref{mainresult_general})
  has been obtained under a physical assumption of
 the observation point location  ${\bf r}$ to be chosen "far enough" ( much further away than the wavelength at a given energy/frequency) from the point of attached antenna/channel.
 Mathematically this condition has been implemented by considering the value of all scalar products $\left\langle r|\textbf{w}_c\right\rangle$ to be negligible in comparison with
 the norms $\gamma_c=|\textbf{w}_c|^2$ for every $c=1,\ldots,M$.\\[1ex]
{\bf Remark 2}. In the course of derivation it has been also assumed that no irreversible losses of flux occur inside the cavity domain.
 In real microwave experiments this is hardly a realistic assumption, unless resonator walls made of superconducting material, like e.g. in
 \cite{dietz2019partial}. It is however well-known how to account for the uniform absorption in cavity walls in the framework of the Heidelberg approach,
 see e.g. \cite{schafer2003correlation,fyodorov2005scattering}.  The idea is that absorption can be treated as loss of flux in the multitude of
 unobserved open channels, very weakly coupled to the cavity.  From this angle, one can add to $M$ observed channels a big number $\tilde{M}\gg 1$
 of channels numbered by channel indices $c=M+1,\ldots,M+\tilde{M}$, all with the same coupling strength: $g_{M+1}=\ldots =g_{\tilde{M}+M}:=g_a$ and
consider the limits $\tilde{M}\to \infty$ and $g_a\to \infty$ while keeping $\tilde{M}/g_a=\epsilon$ fixed. It is easy to check the result of this procedure amounts to
adding to the integrand,  in expressions like Eq.(\ref{mainresult_general}), an extra  factor $e^{-\epsilon(\lambda_1-\lambda_2)}$.
The dimensionless parameter $\epsilon$ should be then interpreted as the effective rate of absorption.
 An alternative procedure for arriving to the same result amounts to
 adding a small positive imaginary part $i\eta$ to the energy $E$ in the formulation of Heidelberg model, see Eq.(\ref{VWZ}) or its equivalent formulation
 Eq.(\ref{eq:S-matrix form1}). In particular, it is evident from Eq.(\ref{eq:S-matrix form1}) that the change $E\to E+i\eta$ entails the loss of S-matrix unitarity, indicating that such a procedure accounts for the irreversible loss of incoming flux in the cavity due to absorption.
 One then finds that at the level of final formulas the net result is exactly the same factor $e^{-\epsilon(\lambda_1-\lambda_2)}$ in the integrands, with
  the parameter $\epsilon$ given by the ratio of the imaginary part $\eta$ to the mean level spacing $\Delta$. This fact shows the equivalence of the two methods.
  It is easy to see that the additional exponential factor in the integrand immediately converts the most distant tails of the intensity distribution ${\cal P}(I)$ from power law to exponential ones, so that the powerlaw behaviour can be observed only in a finite interval of intensities $1\ll I/{\cal I}\ll \epsilon^{-1}$.\\[1ex]
{\bf Remark 3}. If the incoming flux is nonvanishing only in a single channel $c=1$,
the density (Eq.\ref{singlechanold}) coincides with $M=1$ case of the distribution of the photodissociation cross-section $\sigma(E)$
in quantum chaotic systems  studied in \cite{fyodorov1998photodissociation}, see eq. (18) in that paper. Such coincidence is not at all accidental and is to be expected.
 Namely, in the Heidelberg approach the cross-section
can be represented as $\sigma(E)\propto \mbox{Im} \left\langle m\right|\left(E-\hat{\cal H}_{\text{eff}}\right)^{-1}\left|m\right\rangle$, with $\hat{\cal H}_{\text{eff}}$ defined after Eq.(\ref{VWZ}) and $\left|m\right\rangle$ being a fixed nonrandom vector, related to a dipole moment operator. On the other hand, for $M=1$ one may use the identity
\[
\left|\textbf{w}_{\bf a}\right\rangle \left\langle\textbf{w}_{\bf a}\right|= |a_1|^2 \left|\textbf{w}_1\right\rangle\left\langle\textbf{w}_1\right|
\]
\begin{equation}\label{chansum1}
=\frac{|a_1|^2}{2i}\left(\left(E-\hat{\cal H}_{\text{eff}}\right)-\left(E-\hat{\cal H}_{\text{eff}}\right)^{\dagger}\right),
\end{equation}

which, when substituted to  Eq.(\ref{intenstart}) shows that the local intensity in this case is proportional to the diagonal element of the resolvent:
\begin{equation}
I=|a_1|^2\mbox{Im} \left\langle{\bf r}\right|\left(E-\hat{\cal H}_{\text{eff}}\right)^{-1}\left|{\bf r}\right\rangle.
\end{equation}
 We see that indeed intensity for $M=1$ is statistically equivalent to the photodissociation cross-section $\sigma(E)$, up to a constant proportionality factor.

 If incoming fluxes are nonvanishing in more than one channel, then
 $$\left|\textbf{w}_{\bf a}\right\rangle \left\langle\textbf{w}_{\bf a}\right|=\sum_c |a_c|^2\left|\textbf{w}_c\right\rangle\left\langle\textbf{w}_c\right|+\sum_{c,c'} \overline{a}_{c'}{a_{c}}\left|\textbf{w}_c\right\rangle\left\langle\textbf{w}_{c'}\right|$$
 and is never proportional to
 \begin{equation}\label{chansum1}
 \sum_c \left|\textbf{w}_c\right\rangle\left\langle\textbf{w}_c\right|=\frac{1}{2i}\left(\left(E-\hat{\cal H}_{\text{eff}}\right)-\left(E-\hat{\cal H}_{\text{eff}}\right)^{\dagger}\right).
 \end{equation}
  As the result, the correspondence between the diagonal part of the resolvent  and the local intensity is lost, hence for $M>1$ the distribution in Eq.(\ref{perfect1})  is different from
 the corresponding distribution for the normalized cross-section $q=\frac{\sigma}{\left\langle\sigma\right\rangle}$ at perfect coupling, derived originally in  \cite{misirpashaev1997spontaneous}. The latter can be given  for systems of any symmetry $\beta=1,2,4$ as
 \begin{equation}\label{perfect2}
 {\cal P}_M(q)\propto\frac{q^{\frac{\beta M}{2}-1}}{(1+q)^{\beta M+1}}.
 \end{equation}
Note however that for $\beta=2$ the same tail behaviour is shared by Eq.(\ref{perfect1}) and by Eq.(\ref{perfect2}).
 \\[1ex]
{\bf Remark 4}. Finally, in the case when the waves are fed via a single channel $c=1$ one may consider the limit $g_1\gg g_c, \forall c=2,\ldots,N$
describing the case of extremely weak coupling of the feeding channel. In such a limit the point intensity studied in this paper should coincide, after appropriate normalization, with the so-called "transmitted power'', whose distribution for $\beta=2$  chaotic systems has been recovered in the Heidelberg approach framework in \cite{rozhkov2003statistics}
 by the method of moments. One can indeed check that our Eq. (\ref{mainresult_singlech}) in this limiting case reproduces the distribution found in \cite{rozhkov2003statistics}.

 Following the same logic,  one should expect the   Eq.(\ref{mainresult_singlech})  to be itself deducible as a limiting case from the distribution of the modulus of the off-diagonal element of the scattering matrix found in \cite{kumar2013distribution,nock2014distributions}. We check in the Appendix B that this is indeed the case. Let us however stress that (i) the distribution of intensity in the general case, Eq.(\ref{mainresult_general}), can not be deduced in such a way and (ii)  our computation despite being inspired by \cite{kumar2013distribution,nock2014distributions}, was implemented somewhat differently which helped to arrive to the final results in a rather  economic way.

 \subsection{\small Joint probability distribution of intensities at several points}
Consider now a finite number $L\ll N$ of the observation points at locations  ${\bf r}_1, \ldots {\bf r}_L$, each location being both far enough from each of the $M$ antennae,
as well as from each other. We have found that for the case of ergodic systems with broken time-reversal invariance
the joint probability density of the corresponding intensities ${\cal P}_M^{(L)}\left(I_1,\ldots,I_L\right)$ is very simply related to the previously studied one-point density ${\cal P}_M^{(1)}(I):={\cal P}_M(I)$ via:
\begin{equation}\label{multipoint}
{\cal P}_M^{(L)}(I_1,\ldots,I_L)=(-1)^{L-1}\frac{d^{L-1}}{dI^{L-1}}{\cal P}_M(I)\left.\right|_{I=I_1+\ldots+I_L}.
\end{equation}
With this relation it is then straightforward to calculate the probability density $p_M\left(I_{\Sigma}\right)$ for the sum of the intensities $I_{\Sigma}= I_l+\ldots+I_L$:
 \begin{equation}\label{sum_inten}
p_M\left(I_{\Sigma}\right)=\frac{(-1)^{L-1}}{(L-1)!}\,I_{\Sigma}^{L-1}\frac{d^{L-1}}{dI^{L-1}}{\cal P}_M\left(I_{\Sigma}\right).
\end{equation}
In particular, in the case of perfectly coupled channels the Eqs.(\ref{sum_inten})  and Eq.(\ref{perfect1}) imply together:
\begin{equation}\label{multipointperfect}
p_M\left(I_{\Sigma}\right)=\frac{(L+M)!}{M!(L-1)!}\,\left(1+\frac{{\cal I}}{I_{\Sigma}}\right)^{-L}\frac{{\cal I}^{M+1}}{I_{\Sigma}\left(I_{\Sigma}+{\cal I}\right)^{M+1}}.
\end{equation}
 Introducing the  intensity "per point" $i_{\Sigma}=I_{\Sigma}/L$ one finds that such object has the finite
limiting probability density as  $L\to \infty$:
\begin{equation}\label{multipointperfect}
p_M\left(i_{\Sigma}\right)=\frac{1}{L}\frac{1}{M!}\frac{{\cal I}^{M+1}}{i^{M+2}_{\Sigma}}\,e^{-\left({\cal I} / i_{\Sigma}\right)}, \quad i_{\Sigma}=\lim_{L\to \infty}I_{\Sigma}/L.
\end{equation}
{\bf Remark 5}. Summing up the intensities in Eq.(\ref{intenstart}) over {\it all} $N$ internal points in the cavity and using the completeness relation
$\sum_{\bf r}\left|{\bf r}\right\rangle\left\langle{\bf r}\right|=\hat{\bf 1}_N$ one finds that $\sum_{\bf r}I_{\bf r}={\bf a}^{\dagger}\hat{\cal Q}{\bf a}$, where
\begin{equation}\label{intenstotal}
  \hat{\cal Q} = \hat{W}^{\dagger} \frac{1}{E-\hat{H}-i\hat{W}\hat{W}^{\dagger}}\,\frac{1}{E-\hat{H}+i\hat{W}\hat{W}^{\dagger}}\hat{W}.
\end{equation}

The $M\times M$ matrix $\hat{\cal Q}$ is one of most important objects in scattering theory known as the Wigner-Smith time delay matrix. Various aspects of its statistical properties in wave-chaotic systems enjoyed intensive research over several decades, both in the framework of RMT, see the review  \cite{texier2016wigner} and references therein, as well as by semiclassical methods \cite{kuipers2014efficient,novaes2015statistics,novaes2023semiclassical}.
In particular, for the perfect coupling in all channels the distribution of $\hat{\cal Q}$ is known explicitly for all $\beta=1,2,4$ \cite{brouwer1997quantum}, see also \cite{savin2001reducing,grabsch2018wigner} for non-perfect couplings. Combining that distribution with Eq.(\ref{intenstotal})  it is easy to verify that $\sum_{\bf r}I_{\bf r}/{\cal I}$, with ${\cal I}$ defined in Eq.(\ref{influx}), is distributed in the same way
as the diagonal entries $\hat{\cal Q}_{cc}$ of the matrix $\hat{\cal Q}$. In turn, for the perfect coupling the latter entries are known to be distributed in  the same way as {\it partial} delay times $\tau$ \cite{savin2001reducing} whose probability density is explicitly given by
\begin{equation}\label{partial_perfect}
p_M\left(t\right)=\frac{\left(\frac{\beta }{2}\right)^{\frac{\beta M}{2}+1}}{\Gamma(\frac{\beta M}{2}+1)}\frac{1}{t^{\frac{\beta M}{2}+2}}\,
e^{-\left(\frac{\beta }{2t}\right)},\quad  t=\tau \Delta
\end{equation}
where $\Delta\sim N^{-1}$ is the mean level spacing in the closed cavity.
 We then see that the distribution Eq.(\ref{multipointperfect}) of intensity per point $i_{\Sigma}/{\cal I}$  considered in the limit of many observation points $1\ll L\ll N\to \infty$ coincides for $\beta=2$ with the distribution of the total scaled intensity $\Delta \sum_{\bf r}I_{\bf r}/{\cal I}$, i.e. sampled accross the whole cavity.
 This matching implies that the same result will be valid for the (properly scaled) sum of intensities over $L\sim N^{\epsilon}$ points, for any $0<\epsilon\le 1$.

{\bf Remark 6}.  For systems with preserved time reversal invariance with values
$\beta=1,4$ the problem of finding the full distribution of the local intensity $I_{\bf r}$ for arbitrary coupling constants $g_c, c=1,\ldots,M$
and its further $L$-point generalizations remains largely open, apart from $M=1$ case and $L=N$ case, where distributions of partial time delays are known even
at the crossover between $\beta=1$ and $\beta=2$, see \cite{fyodorov1997parametric}.
However one may safely conjecture that the far tail for all these quantities should be universally given by ${\cal P}(I\gg {\cal I})\sim  I^{-\frac{\beta M}{2}+2}$,
as this behaviour in all cases is expected to be controlled by the density of narrow resonances, see discussion in p.1967 of \cite{Fyodorov1997JMP}.

\subsection{\small Distribution of the maximal and minimal intensities in a multipoint observation}

Having at our disposal the joint probability density ${\cal P}_M^{(L)}(I_1,\ldots,I_L)$ given by Eq.(\ref{multipoint}) one can pose a natural question
of the distribution of the maximal and minimal value in the observed pattern:
\begin{equation}\label{maxmindef}
I_{\text{max}}= \text{max}(I_1,\ldots,I_L), \quad I_{\text{min}}= \text{min}(I_1,\ldots,I_L).
\end{equation}
  Note that extreme values of the intensity field in chaotic reverberation chambers were studied experimentally in \cite{gradoni2010extreme}.

 The joint probability (\ref{multipoint}) implies that intensities in different spatial points are in general correlated, apart from the only case
 when the single-point intensity is given by the exponential Rayleigh law ${\cal P}_M(I)\propto e^{-I/\overline{I}}$.
Thus the posed questions belong to the domain of extreme value statistics of many correlated variables $I_k$, which attracted a lot of  attention  in
recent years, especially  when $L\to \infty$, see \cite{majumdar2020extreme} for a review. One of the most studied cases in this area is one inspired by the pattern of repelling eigenvalues of large random matrices, with correlations induced via the presence of the Vandermonde factor $\prod_{k<l}|I_k-I_l|$ in the associated joint probability density,   ultimately leading to the famous Tracy-Widom distribution for the associated extreme values \cite{majumdar2020extreme}.
To this end it is necessary to stress that the correlations in the pattern of intensities emerging in our problem are of a very different nature, and induced rather by the joint probability density depending on all individual intensities only via their sum $\sum_{k}I_k$. Extreme value statistics for such case was not much studied, though a special case appeared in \cite{lakshminarayan2008extreme}, which in our language would
correspond to the particularly simple choice   ${\cal P}^{(L)}(I_1,\ldots,I_L)\propto \delta(I_1+\ldots+I_L)$, and very recently also in the context of resetting problems in \cite{Biroli2022exactorder}. This motivated us to perform the analysis in our case in some detail.

After some computations explained in detail in Section (\ref{Inten}) one finds the general relation in terms of the single-point density ${\cal P}_M(I)$:
\begin{equation}\label{maxmingeneral}
\text{Prob}\left(I_{\text{max}}<Y\right)= \sum_{l=0}^L(-1)^l\left(\begin{array}{c}L\\ l\end{array}\right)\int_{lY}^{\infty}{\cal P}_M(I)\,dI,
\end{equation}
whereas $\text{Prob}\left(I_{\text{min}}>Y\right)=\int_{LY}^{\infty}{\cal P}_M(I)\,dI$. In particular, for the perfect coupling case one can use the Eq.(\ref{perfect1}) and get
\begin{equation}\label{maxperfectgen1}
\text{Prob}\left(I_{\text{max}}<Y\right)= \sum_{l=0}^L(-1)^l\left(\begin{array}{c}L\\ l\end{array}\right)\frac{1}{\left(1+l\frac{Y}{\cal I}\right)^{M+1}}
\end{equation}
and\begin{equation}
   \text{Prob}\left(I_{\text{min}}>Y\right)=\frac{1}{\left(1+L\frac{Y}{\cal I}\right)^{M+1}}.
\end{equation}
We see that in such a pattern of $L$ observation points the typical minimal intensity scales as $I_{\text{min}}^{\text{typ}}\sim {\cal I}L^{-1}$ and the limiting density of
the variable $\sigma_{\text{min}}=L\frac{I_{\text{min}}}{\cal I}$ is given by $\rho\left(\sigma_{\text{min}}\right)=(M+1)\left(1+\sigma_{\text{min}}\right)^{-(M+2)}$, thus of the same form as the density
of the one-point intensity.

The statistics of $I_{\text{max}}$ is somewhat more interesting. To start with, consider the simplest case of the Rayleigh law ${\cal P}(I)=\frac{1}{\overline{I}}e^{-I/\overline{I}}$   obtained in the limit of many open channels, keeping the incoming flux per channel finite: $\lim\limits_{M\to \infty}{\cal I}/M=\overline{I}<\infty$, see Eq.(\ref{Rayleigh}). In this case it is easy to see that
  \begin{equation}\label{Rayleighmax}
 \text{Prob}\left(I_{\text{max}}<Y\right)=\left(1-e^{-Y/\overline{I}}\right)^L.
 \end{equation}
Setting $Y/\overline{I}=\ln{L}+q$ we then recover in the limit $L\to \infty$ the Gumbel distribution:
\begin{multline}\label{Rayleighmaxlim}
 \text{Prob}\left(I_{\text{max}}<\overline{I}(\ln{L}+q)\right)=\exp \left(-e^{-q}\right)\,
\end{multline}
smoothly interpolating between zero at $q\to -\infty$ and one for  $q\to \infty$. The Gumbel law is one of the classical Extreme Value Statistics (EVS), and is
fully expected here as the intensities $I_l$ at different points are uncorrelated. Note also that the threshold of extreme values  is located to the leading order
sharply at  $I_{\text{max}}/\overline{I}=\ln{L}+o(1)$.

Turning our attention now to the finite number of channels, the Eq.(\ref{maxperfectgen1})
 can be alternatively represented as
\begin{equation}\label{maxperfectgen2}
\text{Prob}\left(I_{\text{max}}<Y\right)=\frac{1}{M!}
\int_0^{\infty}dv\, v^M e^{-v}\left(1-e^{-v\frac{Y}{\cal I}}\right)^L.
\end{equation}

In the figure 3 below we compare the probability density associated with the Eq.(\ref{maxperfectgen2}) to the results of numerically generated intensity pattern in the Heidelberg model
for the simplest case of a single-channel system. The results show a reasonable overall agreement.

\begin{figure}[h!]
\includegraphics[scale=0.25,angle=00]{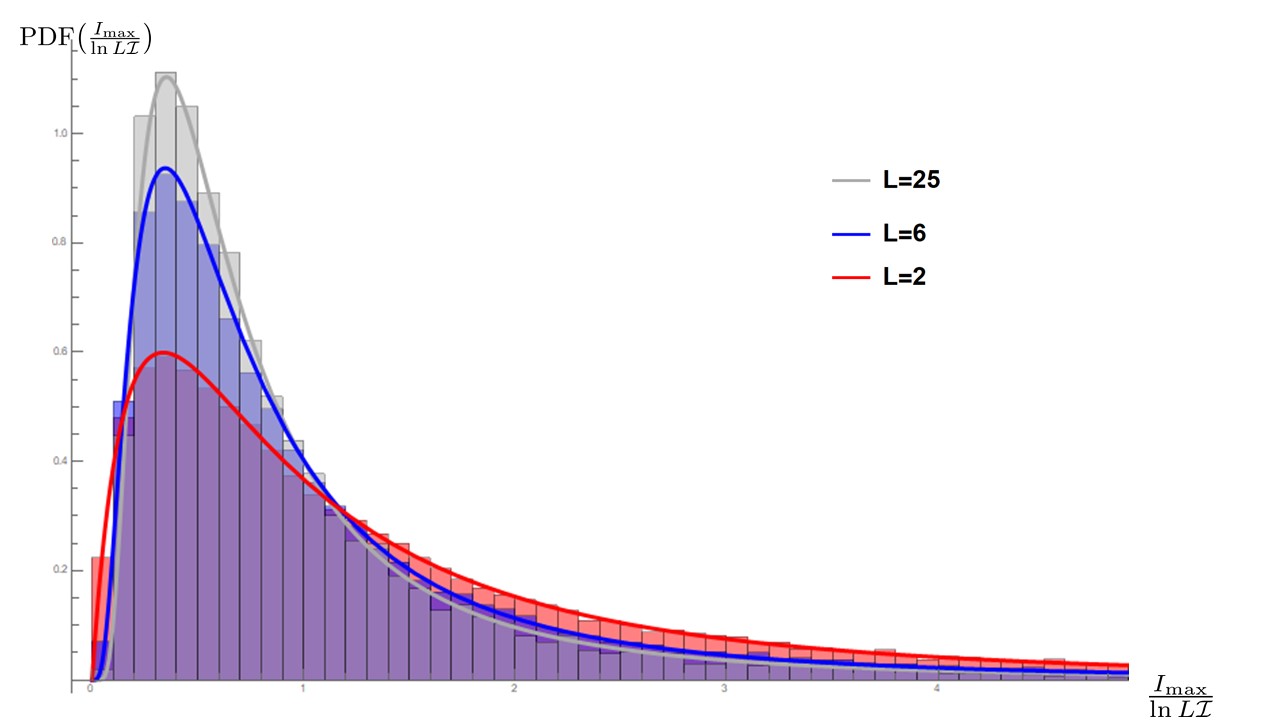}
\caption{\footnotesize The histogram shows numerically evaluated probability density of the maximum intensity in a pattern of $L=2$, $L=6$ and $L=25$ internal points
in the Heidelberg model for a single perfectly open channel system. The intensity data  are first generated according  to Eq.(\ref{intenstart}) at $200$ randomly chosen inner points in a given realization of GUE matrix of size $N=200$. Statistics is built using 100 random subsets of $L$ points per every realization, and
 the displayed data correspond to 1000 different GUE matrix realizations. The solid line for a fixed $L$ is the theoretical prediction given by Eq.(\ref{maxperfectgen2}).}
\end{figure}

Setting in Eq.(\ref{maxperfectgen2})  $Y=\sigma_{\text{max}}{\cal I}\ln{L}$ and considering $\sigma_{\text{max}}>0$ fixed as $L\to \infty$ one first notices that
\[
\lim_{L\to \infty}\left(1-e^{-\sigma_{\text{max}}v\ln{L}}\right)^L=\left\{\begin{array}{cc} 0, & \mbox{if} \,\, 0<v<\sigma_{\text{max}}^{-1}\\1, & \mbox{if} \,\, v>\sigma_{\text{max}}^{-1}\end{array}
\right..
\]
It is now straightforward to see that the typical maximal intensity in a pattern of many points is scaled logarithmically with the number $L$ of observation points:  $I_{\text{max}}^{\text{(typ)}}\sim {\cal I}\ln{L}$, and the limiting  distribution for the properly rescaled maximum intensity is given by:
\[
\lim\limits_{L\to \infty}\text{Prob}\left(I_{\text{max}}<{\cal I}\ln{L}\sigma_{\text{max}}\right)
\]
\begin{equation}\label{maxperectlim}
=\frac{1}{M!}
\int_0^{\sigma_{max}}\, \frac{dt}{t^{M+2}} \exp\left(-\frac{1}{t}\right),
\end{equation}
implying that the probability density for the scaled maximal intensity  $\sigma_{\text{max}}=\frac{I_{\text{max}}}{{\cal I}\ln{L}}$  converges
 in the limit of many observation points to
\begin{equation}\label{maxdenlim}\rho\left(\sigma_{\text{max}}\right)=
\frac{1}{M!{\sigma_{\text{max}}}^{M+2}} \exp{-\frac{1}{\sigma_{\text{max}}}},
\end{equation}
as long as cavity with broken time-reversal invariance is perfectly coupled to antennas.
This type of extreme value statistics resembles the Frechet law density $\rho\left(\sigma\right)=\alpha \sigma^{-\alpha-1}e^{-\sigma^{-\alpha}}$  arising in random patterns of independent, identically distributed random variables $I_k$, each distributed with a powerlaw density $P(I)\sim I^{-(\alpha+1)}$.
Comparing with Eq.(\ref{perfect1}) we may identify $\alpha=M+1$, and  see that had the intensities be independent, the associated Frechet density for the maximum would be
different from Eq.(\ref{maxdenlim}), replacing $\sigma^{-1}$ with $\sigma^{-(M+1)}$  in the exponential factor.
 Eq.(\ref{maxdenlim}) does not seem to appear in the literature on extreme values before.
 Note that it is the same law as the limiting intensity per point, eq.(\ref{multipointperfect}), or partial delay times, eq.(\ref{partial_perfect}).
 An interesting open question is to verify if this property still holds for systems with preserved time-reversal invariance.

To make contact with the previously considered Rayleigh case, one may consider $M\to \infty$ limit in Eq.(\ref{maxperectlim}).
Recalling that in this limit we assume ${\cal I}=M\overline{I}$ we further introduce $\sigma_{\text{max}}M=\overline{\sigma_{\text{max}}}$ and assume it to remain finite
in the limit $M\to \infty$. We also rescale $t=\tau/M$ which gives:
\[
\lim_{M\to \infty}\lim_{L\to \infty}\text{Prob}\left(I_{\text{max}}<\overline{\sigma_{\text{max}}}\,\overline{I}\,\ln{L}\right)
\]
\begin{equation}\label{maxperectlimlim}
=\lim_{M\to \infty}\frac{M^{M+1}}{M!}
\int_0^{\overline{\sigma_{\text{max}}}}\, d\tau e^{-M\left(\ln{\tau}+\frac{1}{\tau}\right)},
\end{equation}
which upon evaluating the integral by the Laplace method yields
\[
\lim_{M\to \infty}\lim_{L\to \infty}\text{Prob}\left(I_{\text{max}}<\overline{\sigma_{\text{max}}}\,\overline{I}\,\ln{L}\right)
\]
\begin{equation}\label{maxperectlimlim1}
=\left\{\begin{array}{cc} 1 & \,\,\mbox{if}\,\,\overline{\sigma_{\text{max}}}>1 \\1/2 & \,\,\mbox{if}\,\,\overline{\sigma_{\text{max}}}=1 \\
0 & \,\,\mbox{if}\,\,\overline{\sigma_{\text{max}}}<1\end{array} \right.\,.
\end{equation}
This agrees with the fact that the threshold of extreme values in this case is sharply at $I_{\text{max}}=\overline{I}\,\ln{L}$, but such interchange of limits (first $L\to \infty$, then $M\to \infty$) misses the fine-scale Gumbel distribution, replacing it by the step function. To improve on that one has to consider the following double scaling limit in Eq.(\ref{maxperfectgen2}): both $M$ and $L$ tending to infinity in such a way that $\frac{\ln{L}}{\sqrt{M}}=c$, with $c\in[0,\infty)$ kept constant, and
also $\lim\limits_{M\to \infty}{\cal I}/M=\overline{I}<\infty$. In such a limit one finds that
\[
   \lim_{L\to \infty}\text{Prob}\left(I_{\text{max}}<\overline{I}\left(\ln{L}+q\right)\right)
\]
\begin{equation}\label{doublescalinglim}
   =\int_{-\infty}^{\infty}\frac{dw}{\sqrt{2\pi}}\,e^{-\frac{w^2}{2}}\exp\left(-e^{-q-cw}\right),
\end{equation}
providing a family of interpolating distributions and reducing to Gumbel for $c=0$.

\section{Outline of the method and derivations of the main results.}

\subsection{Distribution of one-point intensity}
To characterize the distribution of one-point intensity $I_{\bf r}$ we use the method of Laplace transform generating functions, and
aim to calculate for $p>0$ the function
\begin{equation}\label{Laplace1}
{\cal L}(p):=\left\langle e^{-pI_{\bf r}}\right\rangle=\left\langle e^{-pu^*({\bf r})u({\bf r})}\right\rangle,
\end{equation}
where  $u({\bf r}):=\left\langle{\bf r}|{\bf u}\right\rangle$ is the amplitude of the wave in a point ${\bf r}$ inside the cavity,
and angular brackets stand for averaging performed over the random matrix Hamiltonian $\hat{H}$, assumed to be represented by a GUE matrix.

As the first step of the evaluation we find it to be expedient to use a variant of the Gaussian (Hubbard-Stratonovich) transformation,
representing ${\cal L}(p)$ as
\begin{equation}\label{H-s laplace transformA}
\left\langle e^{-pu^*({\bf r})u({\bf r})}\right\rangle=\int\frac{dq^{*}dq}{\pi}e^{-q^{*}q} {\cal R}(q,q^*),
\end{equation}
where we defined
\begin{equation}\label{H-s laplace transformA}
 {\cal R}(q,q^*):=\left\langle e^{-i\sqrt{p}\left(q^{*}u({\bf r})+qu^*({\bf r})\right)}\right\rangle\,,
\end{equation}
which by using  Eq.(\ref{udef}) below can be equivalently written as
\[
 {\cal R}(q,q^*)=
 \]
 \begin{equation}\label{RR0}
=\left\langle e^{-i\sqrt{p}\left(q^{*}\left\langle{\bf r}\right| \frac{1}{E-\hat{H}+i\hat{W}\hat{W}^{\dagger}}\left|\textbf{w}_{\bf a}\right\rangle+q\left\langle{\bf w}_{\bf a}\right| \frac{1}{E-\hat{H}+i\hat{W}\hat{W}^{\dagger}}\left|{\bf r}\right\rangle
 \right)}\right\rangle.
\end{equation}
  We also recall a more general Gaussian identity
\begin{equation}\label{standard Gaussian}
\int e^{-\boldsymbol{z}^{\dagger}\hat{A}\boldsymbol{z}-\left(\boldsymbol{a}^{\dagger}{\bf z}+\boldsymbol{z}^{\dagger}{\bf b}\right)}d\boldsymbol{z}d\boldsymbol{z}^{\dagger}=\frac{\pi^{N}}{\det \hat{A}}\exp\left(\boldsymbol{a^{\dagger}}\hat{A}^{-1}\boldsymbol{b}\right)
\end{equation}
valid, as long as the integral over $ \boldsymbol{z}\in \mathbb{C}^N$ is convergent, for any  $N\times N$ matrix $\hat{A}$ and any complex-valued $N-$component vectors $\boldsymbol{a}, \boldsymbol{b}$.

Let us recall the definition:
\begin{equation} \label{udef}
u({\bf r}):=\left\langle{\bf r}\right| \frac{1}{E{\bf 1}_N-\hat{H}+i\hat{\Gamma}_{\eta}}\left|\textbf{w}_{\bf a}\right\rangle,
\end{equation}
where we defined
\begin{equation}
\hat{\Gamma}_{\eta}=\eta{\bf 1}_N+\pi\sum_{c=1}^{M}{\bf w}_{c}\otimes{\bf w}_{c}^{\dagger},\quad\epsilon>0,
\end{equation}
with $\eta>0$ being a regularization parameter, physically chosen to be proportional to the uniform absorption in the sample.

Now we use Eq.(\ref{standard Gaussian}) for $\hat{A}=-i\left(E-\hat{H}+i\hat{\Gamma}_{\eta}\right)$ to observe that
\begin{equation}\label{HS1}
e^{-i\sqrt{p}q^{*}u({\bf r})}\propto \det\left(E{\bf 1}_N-\hat{H}+i\hat{\Gamma}_{\eta}\right)
\end{equation}
\[
\times \int\, d\boldsymbol{z}_1d\boldsymbol{z}_1^{\dagger}\,e^{i\boldsymbol{z}_1^{\dagger}\left(E{\bf 1}_N-\hat{H}+i\hat{\Gamma}_{\eta}\right)\boldsymbol{z}_1-ip^{1/4}\left(q^*\left\langle{\bf r}|\boldsymbol{z}_1\right\rangle+\left\langle\boldsymbol{z}_1|{\bf w}_{\bf a}\right\rangle\right)}.
\]
Similarly, for $\hat{A}=i\left(E-\hat{H} - i\hat{\Gamma}_{\eta}\right)$ we may see that
\begin{equation}\label{HS2}
e^{-i\sqrt{p}qu^*({\bf r})}\propto \det{\left(E{\bf 1}_N-\hat{H}-i\hat{\Gamma}_{\eta}\right)}
\end{equation}
\[
\times \int\, d\boldsymbol{z}_2d\boldsymbol{z}_2^{\dagger}\, e^{-i\boldsymbol{z}_2^{\dagger}\left(E{\bf 1}_N-\hat{H}-i\hat{\Gamma}_{\eta}\right)\boldsymbol{z}_2+ip^{1/4}\left(q\left\langle{\bf w}_{\bf a}|\boldsymbol{z}_2\right\rangle-\left\langle\boldsymbol{z}_2|\boldsymbol{r}\right\rangle\right)}.
\]
Note that here and below we systematically disregard the proportionality constants, restoring them in final expressions by normalization conditions,
and also find it convenient intermittently use the bra-ket notations for the scalar products, e.g.
$\left\langle\boldsymbol{z}|{\bf w}_{\bf a}\right\rangle\equiv \boldsymbol{z}^{\dagger}{\bf w}_{\bf a}$. Another useful remark is that there is a certain freedom in choosing the arrangement of the variables $q,q^*$ in front of scalar products in the exponents, and we exploited it in two different ways in Eq.(\ref{HS1} and Eq.(\ref{HS2}).
This choice will be {\it aposteriori} justified by very essential simplification of the forthcoming calculations.

On the other hand, the determinant factors entering Eq.(\ref{HS1} and Eq.(\ref{HS2}) can be represented as Gaussian integrals over anticommuting $N-$ vectors $\boldsymbol{\chi}_{\sigma}$ and $\boldsymbol{\chi}_{\sigma}^{*}$
with $\sigma=1,2$:
\[
\det\left(E-\hat{H} \pm \hat{\Gamma}_{\eta}\right)
\]
\[ =\int d\boldsymbol{\chi}_{\sigma}d\boldsymbol{\chi}_{\sigma}^{*}\exp\left(-i\boldsymbol{\chi}_{\sigma}^{\dagger}\left[E-\hat{H} \pm \hat{\Gamma}_{\eta}\right]\boldsymbol{\chi}_{\sigma}\right)
\]
with no issues of convergence arising in this case by definition.

It is convenient to combine vectors with commuting and anticommuting components in a single $4N$-dimensional supervector $\boldsymbol{\Phi}$ defined as
\begin{equation}
\boldsymbol{\Phi}=\left(\begin{array}{c}
\boldsymbol{z}_1\\
\boldsymbol{\chi}_1\\
\boldsymbol{z}_2\\
\boldsymbol{\chi}_2
\end{array}\right),\quad d\boldsymbol{\Phi}d\boldsymbol{\Phi}^{\dagger}=
d\boldsymbol{z}_{1}\boldsymbol{dz}_{1}^{\dagger}d\boldsymbol{\chi}_{1}d\boldsymbol{\chi}_{1}^{\dagger}
d\boldsymbol{z}_{2}d\boldsymbol{z}_{2}^{\dagger}d\boldsymbol{\chi}_{2}d\boldsymbol{\chi}_{2}^{\dagger},
\end{equation}
and also introduce supermatrices  $\hat{L}=\text{diag}\left(1,1,-1,1\right),\quad \hat{\Lambda}=\text{diag}\left(1,1,-1,-1\right)$.
To shorten the notation we in most cases do not distinguish between the number $1$ and identity matrix ${\bf 1}_N$ when the dimensions
are evident from the context.

As the result, we can rewrite the function ${\cal R}(q,q^*)$ in Eq.(\ref{RR0}) as
\[
 {\cal R}(q,q^*)=\iint d\boldsymbol{\Phi}d\boldsymbol{\Phi}^{\dagger}\left\langle e^{i\boldsymbol{\Phi}^{\dagger}\left((E-\hat{H})\hat{L}+i\hat{L}\hat{\Lambda}\hat{\Gamma}_{\eta}\right)\boldsymbol{\Phi}}\right\rangle
 \]
 \begin{equation}
 \times e^{-ip^{1/4}\left(\boldsymbol{\Phi}^{\dagger}\boldsymbol{\xi}_1+\boldsymbol{\xi}_2^{\dagger}\boldsymbol{\Phi}\right)},
\end{equation}
where the supervectors $\boldsymbol{\xi}_{\sigma}$ are given by
\begin{equation}\label{xidef}
\boldsymbol{\xi}_{1}=\left(\begin{array}{c}
{\bf w}_{\bf a}\\
0_N\\
\boldsymbol{r}\\
0_N
\end{array}\right),\quad \boldsymbol{\xi}^{\dagger}_{2}=\left(q^*\boldsymbol{r}^\dagger, 0_N^T, -q\,{\bf w}^{\dagger}_{\bf a},0_N^T\right)\,.
\end{equation}

Closely following a variant of the supersymmetry approach as exposed e.g. in \cite{Fyodorov1997JMP} (one may consult also the lectures \cite{Mirlin2000} for the detail of similar procedures) one can  perform the average over GUE matrices $\hat{H}$ and, after exploiting a supermatrix version of the Hubbard-Stratonovich transformation and peforming the Gaussian integrals over   the supervectors $\boldsymbol{\Phi}$, arrive at the following representation in terms of a $4\times 4$ supermatrix $\hat{R}$ (see the above references for its structure motivated by convergence arguments):
\[
{\cal R}(q,q^*)=
\]
\begin{equation}\label{Rint}
=\int d\hat{R}\,e^{-\frac{N}{2}\text{Str}\hat{R}^{2}} \text{Sdet}^{-1}\left(\left(1_{N}\otimes \hat{L}^{1/2}\right)\hat{G}\left(1_{N}\otimes
\hat{L}^{1/2}\right)\right)
\end{equation}
\[
\times e^{-p^{1/2}\boldsymbol{\xi}_2^{\dagger}\left(1_{N}\otimes \hat{L}^{1/2}\right)\hat{G}\left(1_{N}\otimes \hat{L}^{1/2}\right)\boldsymbol{\xi}_1},
\]
where we introduced the $4N$ component supermatrix $\hat{G}=-i1_{N}\otimes (E-\hat{R})+\hat{\Gamma}_{\eta}\otimes\hat{\Lambda}$.
In what follows we assume the scaling $\eta=\epsilon/2N$, with $\epsilon$ fixed as $N\to \infty$, and in the limit $N\gg 1$ perform the $\hat{R}-$ integral in Eq.(\ref{Rint})
by the saddle-point method, assuming the number of channels $M$ being fixed. Repeating the same steps as in  \cite{Fyodorov1997JMP}, the $\hat{R}-$ integral is reduced to one over a saddle-point manifold parametrized by a $4\times 4$ supermatrix $\hat{Q}=\hat{T}\hat{\Lambda}\hat{T}^{-1}$ where
supermatrices $\hat{T}$ satisfy $\hat{T}^{\dagger}\hat{L}\hat{T}=\hat{L}$. In the Appendix A we give an explicit parametrization of these matrices for convenience of the reader.

To simplify the presentation we also assume for simplicity $E=0$, the results for general $E$ are obtained via the well-known rescaling using the semicircular density of GUE eigenvalues as in \cite{Fyodorov1997JMP}. After all these steps one arrives at the following representation:
\[
{\cal R}(q,q^*)=\int d\mu(\hat{Q})\,e^{-\frac{1}{2}\epsilon\text{Str}\hat{Q}\hat{\Lambda}}\prod_{c=1}^{M}\text{Sdet}\left(1_{4}+\gamma_{c}\left(\hat{\Lambda} \hat{Q}\right)\right)
\]
\begin{equation}\label{QintB}
\times \exp\left(-p^{1/2}\boldsymbol{\xi}_2^{\dagger}\left(\sum_{k=0}^{\infty}\left(-\hat{\Gamma}\right)^{k}\otimes\hat{\tau_k}\right)\boldsymbol{\xi}_{1}\right)
\end{equation}
where we introduced the short-hand notation $\hat{\tau}_{k}\equiv\hat{L}^{-1/2}\hat{Q}\left(\hat{\Lambda} \hat{Q}\right)^{k}\hat{L}^{-1/2}$ and used the parameters $\gamma_c$ as defined in Eq.(\ref{orthchan}).

To evaluate the expression in the exponent of Eq.(\ref{QintB})  we use the definition Eq.(\ref{xidef}) of supervectors $\boldsymbol{\xi}_{1,2}$
to write the $k_{\text{th}}$ term in the sum as
\[
\boldsymbol{\xi}_2^{\dagger}\left(\hat{\Gamma}^{k}\otimes\hat{\tau}_{k}\right)\boldsymbol{\xi}_{1}=q^{*}\left\langle{\bf r}\right|\hat{\Gamma}^{k}\left|{\bf w}_{\bf a}\right\rangle\left(\hat{\tau}_{k}\right)_{b1b1}
\]
\begin{equation}
-q\left\langle{\bf w}_{\bf a}\right|\hat{\Gamma}^{k}\left|{\bf w}_{\bf a}\right\rangle\left(\hat{\tau}_{k}\right)_{b2b1}
\end{equation}
\[
+q^{*}\left\langle\boldsymbol{r}\right|\hat{\Gamma}^{k}\left|\boldsymbol{r}\right\rangle\left(\hat{\tau}_{k}\right)_{b1b2}-q\left\langle{\bf w}_{\bf a}\right|\hat{\Gamma}^{k}\left|\boldsymbol{r}\right\rangle\left(\hat{\tau}_{k}\right)_{b2b2}.
\]
Due to the condition of orthogonality of channels, see Eq.(\ref{orthchan}), we have $\hat{\Gamma}^{k}\left|{\bf w}_{c}\right\rangle=\gamma_{c}^{k}\left|{\bf w}_{c}\right\rangle$ and
\[
\hat{\Gamma}\left|{\bf w}_{\bf a}\right\rangle=\left(\sum_{c=1}^{M}{\bf w}_{c}\otimes {\bf w}_{c}^{\dagger}\right)\left|{\bf w}_{\bf a}\right\rangle=\sum_{c=1}^{M}a_c\gamma_{c}\left|{\bf w}_{c}\right\rangle
\]
 iterating which implies $\hat{\Gamma}^{k}\left|{\bf w}_{\bf a}\right\rangle=\sum_{c=1}^{M}a_c\gamma^k_{c}\left|{\bf w}_{c}\right\rangle$, hence
\begin{equation}
\left\langle{\bf w}_{\bf a}\right|\hat{\Gamma}^{k}\left|{\bf w}_{\bf a}\right\rangle=\sum_{c}^{M}|a_{c}|^2\gamma_c^{k+1}.
\end{equation}
Next important assumption is to consider only the observation of intensity in points far from the channel entrances.
Such a condition is taken into account assuming that $\left\langle{\bf w}_c|{\bf r}\right\rangle=0, \, \forall c=1,\ldots,M$, implying that $\left\langle\boldsymbol{r}\right|\hat{\Gamma}^{k}\left|\boldsymbol{r}\right\rangle=\left\langle{\bf r}|{\bf r}\right\rangle\delta_{k,0}$. In principle, here one may put $\left\langle{\bf r}|{\bf r}\right\rangle=1$, but we leave it in this form as it will help to understand some arising structures later on.

Further using the identity $$\sum_{k=0}^{\infty}\left((-1)^{k}\gamma_{c}^{k}\left(\hat{\Lambda}\hat{ Q}\right)^{k}\right)=\left(1+\gamma_{c}\hat{\Lambda}\hat{Q}\right)^{-1}$$
one can obtain
\[
\boldsymbol{\xi}_2^{\dagger}\left(\sum_{k=0}^{\infty}\left(-\hat{\Gamma}\right)^{k}\otimes\hat{\tau}_{k}\right)\boldsymbol{\xi}_{1}=
\]
\begin{equation}\label{QintC}
=
-q\sum_{c=1}^M|a_c|^2\gamma_{c}\hat{D}_{c}(Q)_{b2b1}+q^{*}\left\langle{\bf r}|{\bf r}\right\rangle\left(\hat{L}^{-1/2}\hat{Q}\hat{L}^{-1/2}\right)_{b1b2}
\end{equation}
where we introduced the supermatrices
\begin{equation}
\hat{D}_{c}(\hat{Q})=\hat{L}^{-1/2}\hat{Q}\left(1+\gamma_{c}\hat{\Lambda}\hat{Q}\right)^{-1}\hat{L}^{-1/2} \quad  \forall \,\,c=1,\ldots,M.
\end{equation}
 Substituting Eq.(\ref{QintC}) into Eq.(\ref{QintB}) gives
 \[
{\cal R}(q,q^*)=\int d\mu(Q)\,e^{-\frac{1}{2}\epsilon\text{Str}\hat{Q}\hat{\Lambda}}\prod_{c=1}^{M}\text{Sdet}\left(1_{4}+\gamma_{c}\left(\hat{\Lambda}\hat{Q}\right)\right)
\]
\begin{equation}\label{QintBB}
\times e^{-p^{1/2}\left[-q\sum_{c=1}^M|a_c|^2\gamma_{c}\hat{D}_{c}(\hat{Q})_{b2b1}+q^{*}\left\langle{\bf r}|{\bf r}\right\rangle\left(\hat{L}^{-1/2}\hat{Q}\hat{L}^{-1/2}\right)_{b1b2}\right]}.
\end{equation}
 Substituting further such a ${\cal R}(q,q^*)$ into Eq.(\ref{H-s laplace transformA}) one may notice that the integral over $q,q^*$ can be easily performed resulting in the following Laplace transformed density of the single-point intensity:
\begin{equation}\label{Laplace2}
{\cal L}(p):=\left\langle e^{-pI_{\bf r}}\right\rangle
\end{equation}
\[
=\int d\mu(Q)\,e^{-\frac{1}{2}\epsilon\text{Str}\hat{Q}\hat{\Lambda}}\prod_{c=1}^{M}\text{Sdet}\left(1_{4}+\gamma_{c}\left(\hat{\Lambda}\hat{Q}\right)\right)
\]
\[
\times \exp\left(-p\left\langle{\bf r}|{\bf r}\right\rangle\sum_{c=1}^M|a_c|^2\gamma_c\left(\hat{L}^{-1/2}\hat{Q}\hat{L}^{-1/2}\right)_{b1b2}\,\hat{D}_c\left(\hat{Q}\right)_{b2b1}\right).
\]
We see that the dependence on the Laplace parameter $p$ in Eq.(\ref{Laplace2}) is extremely simple, which is a direct consequence of the specific
choice made by us in Eqs(\ref{HS1})-(\ref{HS2}). This fact allows us to invert the Laplace transform immediately, getting the probability density for the single-point
intensity via
 \[
{\cal P}_M(I):=\left\langle \delta\left(I-I_{\bf r}\right)\right\rangle
\]
\begin{equation}\label{IntensDist_Q}
=\int d\mu(Q)\,e^{-\frac{1}{2}\epsilon\text{Str}\hat{\Lambda}\hat{Q}}\prod_{c=1}^{M}\text{Sdet}\left(1_{4}+\gamma_{c}\left(\hat{\Lambda}\hat{Q}\right)\right)
\end{equation}
\[
\times \delta\left(I-\sum_{c=1}^M|a_c|^2\gamma_c\left(\hat{L}^{-1/2}\hat{Q}\hat{L}^{-1/2}\right)_{b1b2}\,\hat{D}_c\left(\hat{Q}\right)_{b2b1}\right).
\]
where we eventually replaced $\left\langle{\bf r}|{\bf r}\right\rangle=1$.
Explicit evaluation of such an integral is sketched in the Appendix A, and leads to the form featuring in Eq.(\ref{mainresultA}).

\subsection{Joint probability of intensities at $L$ observation points and extreme value statistics}\label{Inten}

Let us now consider the computation of the joint probability density ${\cal P}_M^{(L)}(I_1,\ldots,I_L)$ of wave intensities $I_1=|u({{\bf r}_1})|^2$, ..., $I_L=|u({{\bf r}_L})|^2$,
where  $u({\bf r}_{l}):=\left\langle{\bf r}_{l}|{\bf u}\right\rangle$ is the amplitude of the wave in a point ${\bf r}_l$ inside the cavity, $l=1,\ldots,L$.
  To start with, we define for the parameters $p_1>0, \ldots,p_L>0$ the joint Laplace transform:
\begin{equation}\label{LaplaceJoint}
{\cal L}(p_1,\ldots,p_L):=\left\langle e^{-\sum_{l=1}^Lp_lu^*({\bf r}_1)u({\bf r}_l)}\right\rangle,
\end{equation}
which after applying Gaussian (Hubbard-Stratonovich) transformations $L$ times takes the form
\begin{equation}\label{H-s laplace transform_joint}
{\cal L}(p_1,\ldots,p_L)=\int\frac{\prod_{l=1}^Ldq_l^{*}dq_l}{(\pi)^L}e^{- \sum\limits_{l=1}^L q_l^{*}q_l} {\cal R}(\{q_l,q_l^*\}),
\end{equation}
where
\begin{equation}\label{RR1}
 {\cal R}(q_1,q_1^*,\ldots,q_L,q_L^*):=\left\langle e^{-i\sum_{l=1}^L\sqrt{p_l}\left(q^{*}_lu({\bf r}_l)+q_lu^*({\bf r}_l)\right)}\right\rangle.
\end{equation}
Now one may notice that Eq.(\ref{innpartmain}) implies
\begin{equation}\label{RR2a}
\sum_{l=1}^L\sqrt{p_l}\,q_l^{*}u({\bf r}_l)=\left\langle{\bf X}\right|\frac{1}{E-\hat{H}+i\hat{W}\hat{W}^{\dagger}}\left|\textbf{w}_{\bf a}\right\rangle,
\end{equation}
where we defined
\begin{equation}\label{RR2aa}
\quad \left\langle{\bf X}\right|=\sum_{l=1}^L\sqrt{p_l}\,q_l^{*}\left\langle{\bf r_l}\right|\,,\quad \left|{\bf X}\right\rangle=\sum_{l=1}^L\sqrt{p_l}\,q_l\left|{\bf r_l}\right\rangle
\end{equation}
and similarly
\begin{equation}\label{RR2a}
\sum_{l=1}^L\sqrt{p_l}\,u^*({\bf r}_l)=\left\langle\textbf{w}_{\bf a}\right|\frac{1}{E-\hat{H}+i\hat{W}\hat{W}^{\dagger}}\left|{\bf X}\right\rangle\,.
\end{equation}
Using the above one can see that we need to evaluate
\[
 {\cal R}(q_1,q_1^*,\ldots,q_L,q_L^*)=
 \]
 \begin{equation}\label{RR3}
 =\left\langle e^{-i\left\langle{\bf X}\right| \frac{1}{E-\hat{H}+i\hat{W}\hat{W}^{\dagger}}\left|\textbf{w}_{\bf a}\right\rangle-i\left\langle\textbf{w}_{\bf a}\right| \frac{1}{E-\hat{H}+i\hat{W}\hat{W}^{\dagger}}\left|{\bf X}\right\rangle}\right\rangle.
\end{equation}
Now, comparing Eq.(\ref{RR3})with Eq.(\ref{RR0}) one may notice that putting $p=q=1$ in the later, and replacing also $\left|{\bf r}\right\rangle \to \left| {\bf X}\right\rangle$ makes the two
expressions identical. Moreover, assuming that all observation points to be located far from every channel entrance implies that the vector $\left|{\bf X}\right\rangle$, being a linear combination of $\left|{\bf r}_l\right\rangle$,  will be orthogonal to all channel vectors $\left|{\textbf{w}_{c}}\right\rangle$. Therefore the evaluation of ensemble average in Eq.(\ref{RR3})
should be simply read off from the expression Eqs.(\ref{QintBB}) for ${\cal R}(q_1,q_1^*)$ implying that
\[
{\cal R}(q_1,q_1^*,\ldots,q_L,q_L^*):=\tilde{{\cal R}}\left(\left\langle{\bf X}|{\bf X}\right\rangle\right)
\]
 \begin{equation}\label{QintAAA}
=\int d\mu(\hat{Q})\,e^{-\frac{1}{2}\epsilon\text{Str}\hat{Q}\hat{\Lambda}}\prod_{c=1}^{M}\text{Sdet}\left(1_{4}+\gamma_{c}\left(\hat{\Lambda}\hat{Q}\right)\right)
\end{equation}
\[
\times e^{- \left\langle{\bf X}|{\bf X}\right\rangle \left(\hat{L}^{-1/2}\hat{Q}\hat{L}^{-1/2}\right)_{b1b2}+\sum\limits_{c=1}^M|a_c|^2\gamma_{c}\hat{D}_{c}(\hat{Q})_{b2b1}},
\]
where we made explicit the fact that ${\cal R}$ depends on the variables $q_l,q_l^*$  for all $l=1,\ldots,L$ (as well as on the Laplace parameters $p_l$)  only via the norm:
\begin{equation}\label{norm X}
\left\langle{\bf X}|{\bf X}\right\rangle=\sum_{l=1}^Lp_lq_l^*q_l,
\end{equation}
where in the above we exploited the inner basis orthogonality: $\left\langle{\bf r}_{l_1}|{\bf r}_{l_2}\right\rangle=\delta_{l_1l_2}$.  Substituting such dependence back into Eq.(\ref{H-s laplace transform_joint})
, passing to polar coordinates: $q_l=\sqrt{R_l}e^{i\theta_l}$, and finally rescaling $R_l\to p_l^{-1}R_l$ leads to:
  \begin{equation}\label{Laplace transform_joint_fin}
{\cal L}(p_1,\ldots,p_L)=\int\tilde{{\cal R}}\left(\sum_{l=1}^LR_l\right)\prod_{l=1}^L\frac{e^{-\frac{R_l}{p_l}}\,dR_l}{p_l}.
\end{equation}
  In such a form the joint Laplace transform can be easily inverted due to the well-known identity involving the Bessel function $J_0(z)$:
 \[
\frac{e^{-\frac{R}{p}}}{p}=\int_0^{\infty}e^{-pI}J_0\left(2\sqrt{IR}\right) dI,
 \]
 yielding the joint probability density of $L$ intensities in the form:
 \begin{equation}\label{jpd_inten_A}
 {\cal P}_M^{(L)}(I_1,\ldots,I_L)=\int_0^{\infty}\tilde{{\cal R}}\left(\sum_{l=1}^LR_l\right)\,\prod_{l=1}^L\,J_0\left(2\sqrt{IR_l}\right)\,dR_l
 \end{equation}
At the next step we use the following chain of identities:
\[
\tilde{{\cal R}}\left(\sum_{l=1}^LR_l\right)=\int_0^{\infty}\tilde{{\cal R}}\left(t\right)\delta\left(t-\sum_{l=1}^LR_l\right)\,dt
\]
\begin{equation}\label{Fourier}
=\int_0^{\infty} dt\tilde{{\cal R}}\left(t\right)
\int_{-\infty}^{\infty}e^{ik\left(t-\sum_{l=1}^LR_l\right)}\frac{dk}{2\pi}.
\end{equation}
Substituting this back to Eq.(\ref{jpd_inten_A}), changing the order of integrations and using that
\begin{equation}\label{LaplBess}
\int_0^{\infty}J_0\left(2\sqrt{IR}\right)e^{-ikR}dR=\frac{1}{ik}e^{\frac{i}{k}I}
\end{equation}
one arrives to the following representation for the joint probability density:
 \begin{equation}\label{jpd_inten_B}
 {\cal P}_M^{(L)}(I_1,\ldots,I_L)=\int_0^{\infty} dt\tilde{{\cal R}}\left(t\right)\Phi_L\left(I_1+\ldots +I_L;t\right),
 \end{equation}
where for the function $\Phi_L\left(I;t\right)$ one easily finds that
\begin{equation}\label{jpd_inten_B}
\begin{split}
\Phi_L\left(I;t\right)\equiv \int_{-\infty}^{\infty}e^{i\left(kt+\frac{I}{k}\right)}\frac{dk}{2\pi (ik)^L} \\ =(-1)^{L-1}\frac{d^{L-1}}{dI^{L-1}}J_0\left(2\sqrt{It}\right).
 \end{split}
 \end{equation}
 Here in the last step we used the inversion of Eq.(\ref{LaplBess}).

This finally implies:
\begin{equation}\label{jpd_inten_B}
 {\cal P}_M^{(L)}(I_1,\ldots,I_L)=(-1)^{L-1}\frac{d^{L-1}}{dI^{L-1}}
 \int_0^{\infty} dt\tilde{{\cal R}}\left(t\right)J_0\left(2\sqrt{It}\right)
 \end{equation}
 \begin{equation}\label{jpd_inten_BB}
 =(-1)^{L-1}\frac{d^{L-1}}{dI^{L-1}}P_{L=1}(I)\left|_{I=\sum_{k=1}^LI_k}\right.,
 \end{equation}
 coinciding with Eq.(\ref{multipoint}).

  To reflect that this joint probability density depends on individual intensities only via their sum $\sum_{k=1}^LI_k$ we define the function $\tilde{{\cal P}}_M^{(L)}(I)$ via
  ${\cal P}_M^{(L)}(I_1,\ldots,I_L)=\tilde{{\cal P}}_M^{(L)}\left(\sum_{k=1}^LI_k\right)$. In particular,
  for finding the probability density for the sum of all intensities, Eq.(\ref{sum_inten}),  we use the identity:
 \[
 \int_0^{\infty} f\left(\sum_{k=1}^LI_k\right)\delta\left(I-\sum_{k=1}^LI_k\right)\prod_{k=1}^LdI_k =\frac{I^{L-1}}{(L-1)!} f(I).
 \]
 Our next step is to consider the simplest extreme value statistics, the distributions of the maximal and the minimal value in the pattern, defined as
 \begin{equation}\label{maxmingeneraldef}
\text{Prob}\left(I_{\text{max}}<Y\right)= \int_{0}^{Y}{\cal P}_M^{(L)}(I_1,\ldots,I_L)\prod_{k=1}^{L}dI_k
\end{equation}
and similarly $$\text{Prob}\left(I_{\text{min}}>Y\right)=\int_{Y}^{\infty}{\cal P}_M^{(L)}(I_1,\ldots,I_L)\prod_{k=1}^{L}dI_k.$$

We will concentrate on the former as the most interesting. Using the same type representation as in Eq.(\ref{Fourier}):
 \begin{equation}\label{rep}
 \tilde{{\cal P}}_M^{(L)}\left(\sum_{l=1}^LI_l\right)=\int_0^{\infty} \tilde{{\cal P}}_M^{(L)}(t)\,dt\int_{-\infty}^{\infty}e^{ik\left(t-\sum_{l=1}^LI_l\right)}\frac{dk}{2\pi}
 \end{equation}
 one easily finds:
\begin{equation}\label{maxdist1}
Prob\left(I_{max}<Y\right)= \int_0^{\infty} \tilde{{\cal P}}_M^{(L)}(t)T_L(t;Y)\,dt
\end{equation}
where we defined
\[
T_L(t;Y):= \int_{-\infty}^{\infty}e^{ikt}\left(\frac{1-e^{-ikY}}{ik}\right)^L\frac{dk}{2\pi}.
\]
 Expanding the binomial and using the identity:
 \[
 \int_{-\infty}^{\infty}\frac{e^{ikt}}{(\beta+ik)^{\nu}}\frac{dk}{2\pi}=\frac{t^{\nu-1}}{\Gamma(\nu)}e^{-\beta t}\theta(t), \quad \beta>0,\nu>0
 \]
 where $\theta(t)=1$ for $t>0$ and zero otherwise, one finds
 \[
  T_L(t;Y)  =\sum_{l=0}^L(-1)^l\left(\begin{array}{c}L\\ l\end{array}\right)
  \frac{(t-lY)^{L-1}}{\Gamma(L)}\theta(t-lY).
  \]
 In particular, one can see that $$T_L(t;Y)=\frac{t^{L-1}}{\Gamma(L)}, \quad 0\le t<Y$$ and $T_L(t;Y)=0$ for $t>LY$. This fact, together with the relation  Eq.(\ref{multipoint}) allows to
 integrate by parts in Eq.(\ref{maxdist1}), which eventually results in the first of relations Eq.(\ref{maxmingeneral}).

\section{Conclusion}
With this work we obtained a pretty complete description of intensity statistics inside irregularly shaped microwave resonator in the quantum chaos regime with broken time-reversal invariance, including multipoint distributions and extreme value statistics. In case of finite number of open channels and no absorption inside all expressions can be, in principle, reduced to elementary functions.   In such a case the one-point intensity is generically powerlaw-distributed, in clear difference with the well-known random Gaussian wave conjecture, cf. Eq.(\ref{Gaumod}),  predicting the exponential Rayleigh law. The latter is only recovered in the very open system limit, while keeping the incoming flux per channel constant. If however uniform losses in the cavity (modelled e.g. by infinite number of weakly coupled hidden channels) are taken into account, the power law  remains only valid in a restricted range of intensities, being  cut exponentially at larger values.  Interestingly, we demonstrated  that the joint probability density of intensities sampled at many points depends only on the sum of individual intensities. We do not have a transparent explanation of such a pattern,  though it may be traced back to the statistical independence of the real and imaginary parts of the complex wavefunctions in a closed cavity, so is definitely expected to hold only for systems with fully broken time reversal invariance.
  Even with such a simple dependence,  the intensities at different spatial points are clearly correlated, unless the system is in the Rayleigh regime. In particular, by extracting the statistics of the highest intensity in an observation pattern of $L$ points explicitly in the perfect coupling regime we were able to demonstrate that the ensuing extreme values distribution for fixed $M$ and  $L\to \infty$, Eq.(\ref{maxdenlim}), differs from the classical extreme value statistics. This provides an example of nontrivial EVS which is potentially accessible in experiments, provided the losses due to absorption can be effectively controlled.  The problem  of characterizing multipoint and extreme value statistics in systems with preserved  time reversal invariance remains currently open.  Calculations in the supersymmetry approach for that case are expected to yield much more cumbersome structures, cf. \cite{kumar2013distribution,nock2014distributions} for the statistics of the modulus of off-diagonal entries of the $S-$matrix. In particular, we expect that the property of the joint distribution of intensities depending only on the sum of individual intensities will be lost in the systems with preserved time-reversal invariance. Such a study presents therefore an interesting challenge for the future research.

 Another possible extension is to consider modifications of intensity statistics by the effects of Anderson localization, which are operative in disordered systems of finite spatial dimension. To this end it is worth mentioning that some aspects of wave intensity statistics inside quasi-1D disordered samples  have been recently under
experimental investigation, see e.g. \cite{cheng2017single,huang2020invariance}. In the framework of the supersymmetry formalism exploited in the present work this would require to go beyond the effectively zero-dimensional limit and combine the 1D nonlinear $\sigma-$ model description of interacting diffusive modes, see \cite{FM1994} and references therein, with the Heidelberg model formalism. For a few examples of recent studies of not dissimilar problems see e.g. \cite{huang2020invariance} and \cite{fyodorov2022resonances}.

\subsection*{Acknowledgements}
Y.V.F. acknowledges financial support from EPSRC Grant EP/V002473/1 ``Random Hessians and Jacobians: theory and applications''
and is grateful to Prof. Gregory Schehr for a discussion on extreme value statistics and bringing the references \cite{lakshminarayan2008extreme,Biroli2022exactorder}
to his attention. E.S. is grateful to Prof. Mikhail Feigelman for discussions of the results which helped to improve the presentation.

 \vspace{1cm}

\appendix{ \bf Appendix A: Parameterization of $\hat{Q}-$supermatrices and related formulas}

\vspace{0.5cm}

We use the same parametrization as in \cite{Mirlin2000}, and describe it below for convenience of the reader.

First one defines two unitary $2\times 2$ supermatrices
\begin{equation}
\hat{U}_{1}=\exp\left(\begin{array}{cc}
0 & -\alpha^{*}\\
\alpha & 0
\end{array}\right),\quad \hat{U}_{2}=\exp i\left(\begin{array}{cc}
0 & -\beta^{*}\\
\beta & 0
\end{array}\right)
\end{equation}
where $\alpha, \alpha^{*}, \beta, \beta^*$ are anticommuting variables. In terms of those the $4\times 4$ supermatrix $\hat{Q}$ is defined as

\begin{equation}
\hat{Q}=\left(\begin{array}{ll}
\hat{U}_{1}\\
 & \hat{U}_{2}
\end{array}\right)\left(\begin{array}{cccc}
\lambda_{1} & 0 & i\mu_{1} & 0\\
0 & \lambda_{2} & 0 & \mu_{2}^{*}\\
i\mu_{1}^{*} & 0 & -\lambda_{1} & 0\\
0 & \mu_{2} & 0 & -\lambda_{2}
\end{array}\right)\left(\begin{array}{cc}
\hat{U}_{1}^{-1}\\
 & \hat{U}_{2}^{-1}
\end{array}\right)
\end{equation}

where

\[
\begin{cases}
1\leq\lambda_{1}<\infty,\quad\mu_{1}=|\mu_{1}|e^{i\phi_{1}},\quad|\mu_{1}|^{2}=\lambda_{1}^{2}-1\\
-1\leq\lambda_{2}\leq1,\quad\mu_{2}=|\mu_{2}|e^{i\phi_{2}},\quad,\quad|\mu_{2}|^{2}=1-\lambda_{2}^{2}
\end{cases}
\]

The measure $d\mu(\hat{Q})$ will take the following form

\begin{equation}
d\mu(\hat{Q})=-\frac{d\lambda_{1}d\lambda_{2}}{(\lambda_{1}-\lambda_{2})^{2}}d\phi_{1}d\phi_{2}d\alpha d\alpha^{*}d\beta d\beta^{*}
\end{equation}

It is immediate to check that in this parameterization $\text{Str}\hat{\Lambda}\hat{Q}$ and $\text{Sdet}\left(1_{4}+\gamma_{c}\hat{\Lambda}\hat{Q}\right)$
take the form

\begin{equation}
    \text{Str} \hat{\Lambda}\hat{Q}=2\left(\lambda_{1}-\lambda_{2}\right),
\end{equation}
\[
\text{Sdet}\left(1_{4}+\gamma_{c}\hat{\Lambda}\hat{Q}\right)=\frac{1+2\gamma_{c}\lambda_{1}+\gamma_{c}^{2}}{1+2\gamma_{c}\lambda_{2}+\gamma_{c}^{2}}
\]
correspondingly. We also need the following combinations:
\[
\hat{D}_{c}\left(\hat{Q}\right)_{b1b2}=-i\left[\frac{i\mu_{1}}{1+2\gamma_{c}\lambda_{1}+\gamma_{c}^{2}}\left(1+\frac{\beta^*\beta}{2}\right)\left(1-\frac{\alpha^*\alpha}{2}\right)\right.
\]
\begin{equation}
\left.+i\alpha^{*}\frac{\mu_{2}^{*}}{1+2\gamma_{c}\lambda_{2}+\gamma_{c}^{2}}\beta\right],
\end{equation}
which can be used to get also $\left(\hat{L}^{-1/2}\hat{Q}\hat{L}^{-1/2}\right)_{b1b2}=\lim\limits_{\gamma_c\to 0}\hat{D}_{c}\left(\hat{Q}\right)_{b1b2}$. Similarly,
\[
\hat{D}_{c}\left(\hat{Q}\right)_{b2b1}=-i\left[\frac{i\mu_{1}^{*}}{1+2\gamma_{c}\lambda_{1}+\gamma_{c}^{2}}\left(1+\frac{\beta^*\beta}{2}\right)\left(1-\frac{\alpha^*\alpha}{2}\right)\right.
\]
\begin{equation}
\left.
+i\beta^{*}\alpha\frac{\mu_{2}}{1+2\gamma_{c}\lambda_{2}+\gamma_{c}^{2}}\right].
\end{equation}

Substituting all this to Eq.(\ref{IntensDist_Q}) gives:

 \begin{equation}\label{IntensDist_Qexplicit}
{\cal P}_M(I)=\int d\mu(\hat{Q})\,e^{-\epsilon(\lambda_1-\lambda_2)}\prod_{c=1}^{M}\left(\frac{\lambda_2+g_c}{\lambda_1+g_c}\right)
\end{equation}
\[
\times \delta\left(I-\left\langle{\bf r}|{\bf r}\right\rangle\sum_{c=1}^M|a_c|^2A_c(\hat{Q})\right),
\]
where
\[
A_c(\hat{Q})=\frac{1}{2}\frac{|\mu_1|^2}{\lambda_1+g_c}(1+\beta^*\beta-\alpha^*\alpha-\beta^*\beta\alpha^*\alpha)
\]
\[+\alpha^*\beta\frac{1}{2}\frac{\mu_1^*\mu_2^*}{\lambda_1+g_c}+\beta^*\alpha\frac{1}{2}\frac{\mu_1\mu_2}{\lambda_2+g_c}-\beta^*\beta\alpha^*\alpha
\frac{1}{2}\frac{|\mu_2|^2}{\lambda_2+g_c}.
\]

Now one may expand the Dirac $\delta$-function into anticommuting variables and perform the corresponding integrals, and then over angular variables $\phi_{1,2}$.
After straightforward algebraic manipulations  one arrives at
\begin{equation}\label{IntensDist_Qexplicit1}
{\cal P}_M(I)=\delta\left(I\right)-\frac{d {\cal F}_M(I)}{dI}+\frac{d^2}{dI^2}\left(I {\cal F}_M(I)\right),
\end{equation}
where  ${\cal F}_M(I)$ will be defined in (\ref{IntensDist_Qexplicit1A}) below. Here we note that as explained in the Appendix of the paper \cite{rozhkov2003statistics}  the so-called "Efetov-Wegner" term $\delta\left(I\right)$ in Eq.(\ref{IntensDist_Qexplicit1}) gets eventually cancelled and can be omitted. The function ${\cal F}_M(I)$ is given explicitly by
\begin{equation}\label{IntensDist_Qexplicit1A}
 {\cal F}_M(I)=\int_{1}^{\infty}\int_{-1}^{1} \frac{d\lambda_1d\lambda_2}{(\lambda_1-\lambda_2)^2}e^{-\epsilon(\lambda_1-\lambda_2)}\prod_{c=1}^{M}\left(\frac{\lambda_2+g_c}{\lambda_1+g_c}\right)
\end{equation}
\[
\times \left(I+\frac{|\mu_2|^2}{2}\sum_{c=1}^M\frac{|a_c|^2}{\lambda_2+g_c}\right) \delta\left(I-\frac{|\mu_1|^2}{2}\sum_{c=1}^M\frac{|a_c|^2}{\lambda_1+g_c}\right).
\]
  After further manipulations using $|\mu_1|^2=\lambda_1^2-1, \,|\mu_2|^2=1-\lambda_2^2 $ and noticing that the $\delta-$functional constraint implies
\[
I=\frac{\lambda_1-1}{2}\sum_{c=1}^M|a_c|^2\frac{\lambda_1+1}{\lambda_1+g_c}
\]
\[
=\frac{\lambda_1-1}{2}\sum_{c=1}^M|a_c|^2\left(1-\frac{g_c-1}{\lambda_1+g_c}\right)
\]
and
\[
\left(I+\frac{|\mu_2|^2}{2}\sum_{c=1}^M\frac{|a_c|^2}{\lambda_2+g_c}\right)\left|_{I=\frac{|\mu_1|^2}{2}\sum_{c=1}^M\frac{|a_c|^2}{\lambda_1+g_c}}\right.
\]
\[=
\frac{\lambda_1-\lambda_2}{2}\sum_{c=1}^M|a_c|^2\frac{\lambda_1\lambda_2+g_c(\lambda_1+\lambda_2)+1}{(\lambda_1+g_c)(\lambda_2+g_c)},
\]
we can bring the Eq.(\ref{IntensDist_Qexplicit1A})  to the form featuring in Eq.(\ref{mainresultA}).

\vspace{1cm}


\appendix{ \bf Appendix B: Relation to Nock et al. \cite{nock2014distributions}}\\[1ex]

The paper  \cite{nock2014distributions} provided the explicit result for the probability density of the modulus $|S_{ab}|:=r$ for the $S-$matrix entry between two different channels $a\ne b$, where without reducing generality one may consider $a=1$ and $b=M$.  For the systems with broken time-reversal invariance
  the probability density $P_r(r)$ for the variable $r$ (normalized in such a way
 that $\int_0^{\infty}P_r(r)\,r\,dr=1$) can be found in Eqs.(60)-(62) of  \cite{nock2014distributions} and is represented in the form:
 \begin{equation}\label{Nock1}
 P_r(r)=\frac{1}{r}\frac{\partial}{\partial r} r\frac{\partial}{\partial r} f(r)
 \end{equation}
 where
 \begin{equation}\label{Nock2}
  f(r)=\frac{1}{2}\frac{(g_1+\lambda_1)^2(g_M+\lambda_1)^2}{(g_1+g_M)\lambda_1^2+2(g_1g_M+1)\lambda_1+(g_1+g_M)}\, {\cal U}(r)
 \end{equation}
 and ${\cal U}(r)$ is given by
 \[
  {\cal U}(r)=\int_{-1}^1\frac{d\lambda_2}{(\lambda_1-\lambda_2)^2}\prod_{c=1}^M\frac{g_c+\lambda_2}{g_c+\lambda_1}
 \]
  \begin{equation}
 \label{Nock2a}
\times \left(\frac{\lambda_1^2-1}{(g_1+\lambda_1)(g_M+\lambda_1)}+
 \frac{1-\lambda_2^2}{(g_1+\lambda_2)(g_M+\lambda_2)}\right),
 \end{equation}
 with $\lambda_1$ for a given $r$ being defined via
 \begin{equation}\label{Nock3}
 \lambda_1=\frac{(g_1+g_M)r^2+\sqrt{(g_1-g_M)^2r^4+4r^2(g_1g_M-1)+4}}{2(1-r^2)}.
 \end{equation}

On the other hand recall that
according to Eq.(\ref{VWZ}) we have
 \begin{equation}
|S_{1M}|^2=4\left|\left\langle\textbf{w}_{1}\right| \frac{1}{E-\hat{H}+i\hat{W}\hat{W}^{\dagger}}\left|\textbf{w}_{M}\right\rangle\right|^2:=r^2,
 \end{equation}
where $\hat{W}\hat{W}^{\dagger}=\sum_{c=1}^M \textbf{w}_c\otimes \textbf{w}_c^{\dagger}$.  Consider now the limit $\gamma_M=|\textbf{w}_{M}|^2\to 0$
while keeping $|\textbf
{w}_c|^2 =\gamma_{c}$  of the order of unity for all $c\ne M$.  Physically this corresponds to almost closing the channel with $c=M$, with the effective coupling $g_M\approx \frac{1}{2|\textbf{w}_{M}|^2}\gg g_c, \, \forall c<M$. It is then easy to see that in such a limit $|S_{1M}|^2\to 0$, whereas the
product  $|S_{1M}|^2 g_M/2$ remains finite and simply proportional to the intensity $I$ at a single point inside the cavity given by Eq.(\ref{intenstart}),
provided we reduce the number of open channels by one and consider the incoming wave amplitudes $a_c$ to be nonzero only for $c=1$.
We therefore can extract the probability distribution of the point intensity in such a case by performing the limit $g_M\to \infty$  and $r^2\to 0$  in  Eq.(\ref{Nock2}) while keeping $r^2 g_M=2I$ and $g_c, \, \forall c<M$ finite.
In such a limiting procedure we get:
\[
\lambda_1\to I+\sqrt{I^2+2g_1I+1},  \quad \prod_{c=1}^M\frac{g_c+\lambda_2}{g_c+\lambda_1}\to \prod_{c=1}^{M-1}\frac{g_c+\lambda_2}{g_c+\lambda_1}
\]
and
\[
\frac{1}{2}\frac{(g_1+\lambda_1)^2(g_M+\lambda_1)^2}{(g_1+g_M)\lambda_1^2+2(g_1g_M+1)\lambda_1+(g_1+g_M)}
\]
\begin{equation}
\approx  \frac{g_M}{2}\frac{(g_1+\lambda_1)^2}{\lambda_1^2+2\lambda_1g_1+1}=\frac{g_M}{2}
\frac{(g_1+\lambda_1)}{2\sqrt{I^2+2g_1I+1}}.
\end{equation}
Further we have
\[
\frac{\lambda_1^2-1}{(g_1+\lambda_1)(g_M+\lambda_1)}+\frac{1-\lambda_2^2}{(g_1+\lambda_2)(g_M+\lambda_2)}
\]
\[
\approx \frac{1}{g_M}\left[\frac{\lambda_1^2-1}{(g_1+\lambda_1)}+\frac{1-\lambda_2^2}{(g_1+\lambda_2)}\right]
\]
\begin{equation}
=\frac{(\lambda_1-\lambda_2)}{g_M}\,\frac{\left[g_1(\lambda_1+\lambda_2)+\lambda_1\lambda_2+1\right]}{(g_1+\lambda_1)(g_1+\lambda_2)}
=\frac{(\lambda_1-\lambda_2)}{g_M}\frac{\left(\tilde{g}_1+\lambda_2\right)}{g_1+\lambda_2},
\end{equation}
where we used the definitions  Eq.(\ref{g1tilde}) and  Eq.(\ref{relations1}):
\[
\tilde{g}_1=\frac{g_1\lambda_1+1}{g_1+\lambda_1}=-I+\sqrt{I^2+2g_1I+1}.
\]
Substituting all these factors back to Eq.(\ref{Nock2})-(\ref{Nock2a}) yields finally:
\[
f(r)\to {\cal F}_{M-1}(I)=\frac{1}{4\sqrt{I^2+2g_1I+1}} \frac{1}{\prod_{c=1}^{M-1}(g_c+\lambda_1)}
\]
\begin{equation}
\times \int_{-1}^1\frac{d\lambda_2}{\lambda_1-\lambda_2}(\tilde{g}_1+\lambda_2)\prod_{c=2}^{M-1}(g_c+\lambda_2)
\end{equation}

which together with
$$
P(r)=\frac{1}{r}\frac{\partial}{\partial r} r\frac{\partial}{\partial r} f(r)\to P(I)=4\frac{\partial}{\partial I} I\frac{\partial}{\partial I}  {\cal F}_{M-1}(I)
$$
 reproduces exactly the pair Eq.(\ref{mainresultA})-Eq.(\ref{mainresult_singlech}) with obvious replacement $M\to M-1$.


%

\end{document}